\documentclass[%
 reprint,
 amsmath,amssymb,
 aps,
 prx,
floatfix,
]{revtex4-2}

\usepackage{graphicx}
\usepackage{dcolumn}
\usepackage{bm}
\usepackage{braket}
\usepackage{xcolor}
\usepackage[normalem]{ulem}
\usepackage{amsmath}
\usepackage{bbm}
\usepackage{upgreek}
\usepackage{csquotes}
\usepackage{cleveref}

\def\dashedrule#1#2#3{{%
\dimen1=#2 \divide\dimen1 by 2
\def\@ruledash{%
\rule{\dimen1}{0pt}%
\rule[0.5ex]{#1}{0.4pt}%
\rule{\dimen1}{0pt}}%
\count1=0
\loop%
\ifnum\count1<#3%
\advance\count1 by 1%
\@ruledash%
\repeat}}

\usepackage{physics}
\usepackage{upgreek}
\usepackage{bbm}

\newcommand{\asciimathunit}[1]{\ensuremath{\,\mathrm{#1}}}

\newcommand{\nm}{\asciimathunit{nm}}

\newcommand{\kHz}{\asciimathunit{kHz}}
\newcommand{\MHz}{\asciimathunit{MHz}}
\newcommand{\GHz}{\asciimathunit{GHz}}
\newcommand{\us}{\ensuremath{\,\upmu \mathrm{s}}}
\newcommand{\ms}{\asciimathunit{ms}}
\newcommand{\s}{\asciimathunit{s}}
\newcommand{\mW}{\asciimathunit{mW}}

\newcommand{\degree}{\ensuremath{^\circ}}

\def \loadeff {\ensuremath{80 \%}} 
 
\def \imgdepth {\ensuremath{8.7 \MHz}}

\def \imgfidelity {$\approx$99.8\%}
\def \imgloss {$0.19(2)\%$}
\def \magicangle {$\approx$17$^\circ$ }
\def \DeltaG {\ensuremath{-5.8 \GHz}}
\def \DeltaM {\ensuremath{+1 \GHz}} 
\def \RabiG {\ensuremath{0.33 \MHz}} 
\def \RabiM {\ensuremath{0.26 \MHz}} 
 
\def \RabiClockSigma {\ensuremath{0.11 \MHz}} 
\def \ramseydepth {\ensuremath{230 \kHz}} 
\def \ramseytrapfreq {\ensuremath{9.5 \kHz}} 
\def \aodpower {\ensuremath{2.4\mW}}

\def \PushOutFidelity {\ensuremath{99.91(1) \%}} 
\def \GroundQubitFidelity {\ensuremath{99.968(3) \%}} 
 
\def \MetastableQubitFidelity {\ensuremath{99.12(4) \%}} 
\def \OpticalQubitFidelity {\ensuremath{99.804(8) \%}}  
\def \OpticalQubitInFidelity {\ensuremath{2.0\times 10^{-3}}} 
\def \OpticalQubitInFidelitySim {\ensuremath{1.8\times 10^{-3}}} 
\def \UpFidelity {\ensuremath{98.6(2) \%}} 
\def \DownFidelity {\ensuremath{99.4(1) \%}}

\def \UpNDprob {\ensuremath{97.4(3) \%}}
\def \DownNDprob {\ensuremath{99.0(2) \%}} 
\def \m {\ensuremath{|m\rangle}}
\def \g {\ensuremath{|g\rangle}}
\def \o {\ensuremath{\textit{o}}}
\def \up {\ensuremath{|0\rangle}}
\def \down {\ensuremath{|1\rangle}}
\def \nbar {\ensuremath{0.05(1)}}

\def \QBsplittingG {\ensuremath{1.1 \kHz}} 
\def \QBsplittingM {\ensuremath{1.7 \kHz}} 
\def \gatelength {\ensuremath{0.5-1.4 \us}} 
\def \MetaDecay {\ensuremath{0.77(8) \mathrm{s}^{-1}}} 
\def \MetaDecoh {\ensuremath{1.8(2) \mathrm{s}^{-1}}} 
\def \MetaError {\ensuremath{0.7(1) \%}} 
\def \groundcohdecay {\ensuremath{12(2) \s}}
\def \metacohdecay {\ensuremath{7.2(6) \s}}
\def \ramseyscattrate {\ensuremath{0.05(1) \mathrm{s}^{-1}}} 
\def \imgscattrate {\ensuremath{2.0(4) \mathrm{s}^{-1}}}

\def \OpticalQubitFidelityLS {\ensuremath{99.53(4) \%}} 
\def \ClockSuppressionError {\ensuremath{0.27(2) \%}} 
\def \MCMAQcontrast {\ensuremath{98.2(6) \%}} 
\def \MCMDQcontrast {\ensuremath{95.5(1.0) \%}}

\def \MCOnoresetnbar {\ensuremath{1.3(2)}} 
\def \MCOresetnbar {\ensuremath{0.26(6)}} 
\def \MCOnoresetAQContrast {\ensuremath{12(1) \%}} 
\def \MCRAQcontrast {\ensuremath{97.7(5) \%}}
 
\def \MCRDQcontrast {\ensuremath{95.2(8) \%}} 
\def \lukinFidelity {\ensuremath{99.5 \%}} 
\def \flopsG {\ensuremath{80}} 
\def \flopsO {\ensuremath{80}} 
\def \flopsM {\ensuremath{30}} 
\def \rbintflucerror {\ensuremath{1\times 10^{-4}}} 
\def \rbgscatterr {\ensuremath{3\times 10^{-5}}} 

\begin{document}

\preprint{APS/123-QED}

\title{Mid-circuit operations using the \textit{omg}-architecture in neutral atom arrays}% Force line breaks with \\

\author{Joanna W. Lis\textsuperscript{1}}
\author{Aruku Senoo\textsuperscript{1}}
\author{William F. McGrew\textsuperscript{1}}
\author{Felix R\"{o}nchen\textsuperscript{2}}
\author{Alec Jenkins\textsuperscript{1}}
\author{Adam M. Kaufman\textsuperscript{1}}
\email{adam.kaufman@colorado.edu}
\affiliation{{\normalfont\textsuperscript{1}}JILA, University of Colorado and National Institute of Standards and Technology,
and Department of Physics, University of Colorado, Boulder, Colorado 80309, USA}%
\affiliation{{\normalfont\textsuperscript{2}}Physikalisches Institut, University of Bonn, Wegelerstrasse 8, 53115 Bonn, Germany}%

\date{\today}

\begin{abstract}
We implement mid-circuit operations in a 48-site array of neutral atoms, enabled by new methods for control of the \textit{omg} (optical-metastable-ground state qubit) architecture present in ${}^{171}$Yb. We demonstrate laser-based control of ground, metastable and optical qubits with average single-qubit fidelities of $\mathcal{F}_{g} = \GroundQubitFidelity$, $\mathcal{F}_{m} = \MetastableQubitFidelity$ and $\mathcal{F}_{o} = \OpticalQubitFidelity$. With state-sensitive shelving between the ground and metastable states, we realize a non-destructive state-detection for $^{171}$Yb, and reinitialize in the ground state with either global control or local feed-forward operations. We use local addressing of the optical clock transition to perform mid-circuit operations, including measurement, spin reset, and motional reset in the form of ground-state cooling. In characterizing mid-circuit measurement on ground-state qubits, we observe raw errors of $1.8(6)\%$ on ancilla qubits and $4.5(1.0)\%$ on data qubits, with the former (latter) uncorrected for $1.0(2)\%$ ($2.0(2)\%$) preparation and measurement error; we observe similar performance for mid-circuit reset operations. The reported realization of the \textit{omg} architecture and mid-circuit operations are door-opening for many tasks in quantum information science, including quantum error-correction, entanglement generation, and metrology. 
\end{abstract}

\maketitle

%\tableofcontents

\section{\label{sec:introduction}Introduction}

Quantum state initialization and measurement are essential tasks in quantum information science~\cite{divincenzo2000physical}. In many NISQ-era experiments, these operations bookend an experiment, e.g. in a finite depth quantum circuit, studying a many-body Hamiltonian for quantum simulation, or integrating a phase from a signal Hamiltonian for metrology~\cite{preskill2018quantum, ludlow2015optical}. However, it can be  desirable that measurements, as well as the dissipative operations that are common to state initialization, are interleaved ``mid-circuit" with coherent operations. This capability is especially powerful when performed in a partial fashion, that is, in a manner which leaves a subset of qubits --- ``data qubits" --- unperturbed by the operations applied to its complement, the ``ancilla qubits". Salient areas where these so-called mid-circuit operations are enabling include quantum error correction, measurement-based quantum computing, metrology, entanglement generation, and quantum simulation, to name a few~\cite{briegel2009measurement,terhal2015quantum,rosenband2013exponential,pezze2020heisenberg, kessler2014heisenberg, skinner2019measurement, tantivasadakarn2021long, Feig2021Holographic,lu2022measurement,foss2023}.

With many developing quantum science platforms --- such as superconducting qubits, trapped ions, and neutral-atom arrays  --- the realization of high-fidelity mid-circuit operations (MCOs) is a key focus~\cite{kjaergaard2020superconducting, bruzewicz2019trapped, kaufman2021quantum}.  Within the superconducting qubits and trapped ion architectures, high fidelity mid-circuit measurement has facilitated state teleportation, quantum error correction, and the generation of topologically-ordered entangled states~\cite{riebe2004deterministic,wan2019quantum, negnevitsky2018repeated,krinner2022realizing, iqbal2023topological,acharya2023suppressing}. 
Meanwhile, mid-circuit qubit reset and reuse can decrease resource overhead and increase circuit fidelity, reducing the number of physical qubits and entangling gates needed to execute algorithms~\cite{hua2022exploiting,decross2022qubit}. In trapped ions, mid-circuit cooling of the motional modes is employed to attain high two-qubit gate fidelities \cite{wan2019quantum,negnevitsky2018repeated} and can aid in correcting bit-flip errors \cite{reiter2017dissipative}, while on the superconducting platform the reset alleviates leakage errors~\cite{mcewen2021removing} and enables implementation of autonomous error-correction~\cite{li2023autonomous}. 

For neutral atom arrays as well as trapped ions, mid-circuit operations can be particularly challenging. This is because operations like fluorescence detection, optical pumping, and laser cooling often involve illuminating atoms with near-resonant light, during which an accidental scattering of a single photon can destroy quantum information stored in internal states. Accordingly, the central goal is to devise methods to protect data qubits during these resonant operations. In trapped-ion systems, mid-circuit operations have been performed using additional atomic species, qubit shuttling, and long-lived shelving states~\cite{riebe2004deterministic, Hume2007High, negnevitsky2018repeated, Gaebler2021Suppression, moses2023race}. 

It is only within the past year that mid-circuit qubit readout has been demonstrated in neutral atoms arrays. In a cavity-based approach, two atoms were transported sequentially into the waist of a cavity for state-detection~\cite{deist2022mid}. In a dual-species scheme, the frequency selectivity of atomic transitions enabled one atomic species to be measured, and the measurement outcome was fed-forward as a phase correction to the second atomic species~\cite{singh2023mid}. Finally, in the third approach, through local light-shifting in an alkali-atom qubit array, a single data qubit was selected and shelved into a state dark to the imaging light, albeit where the field sensitivity and the hyperfine structure were key challenges~\cite{graham2023mid}. 

\begin{figure*}
\begin{centering}
\includegraphics{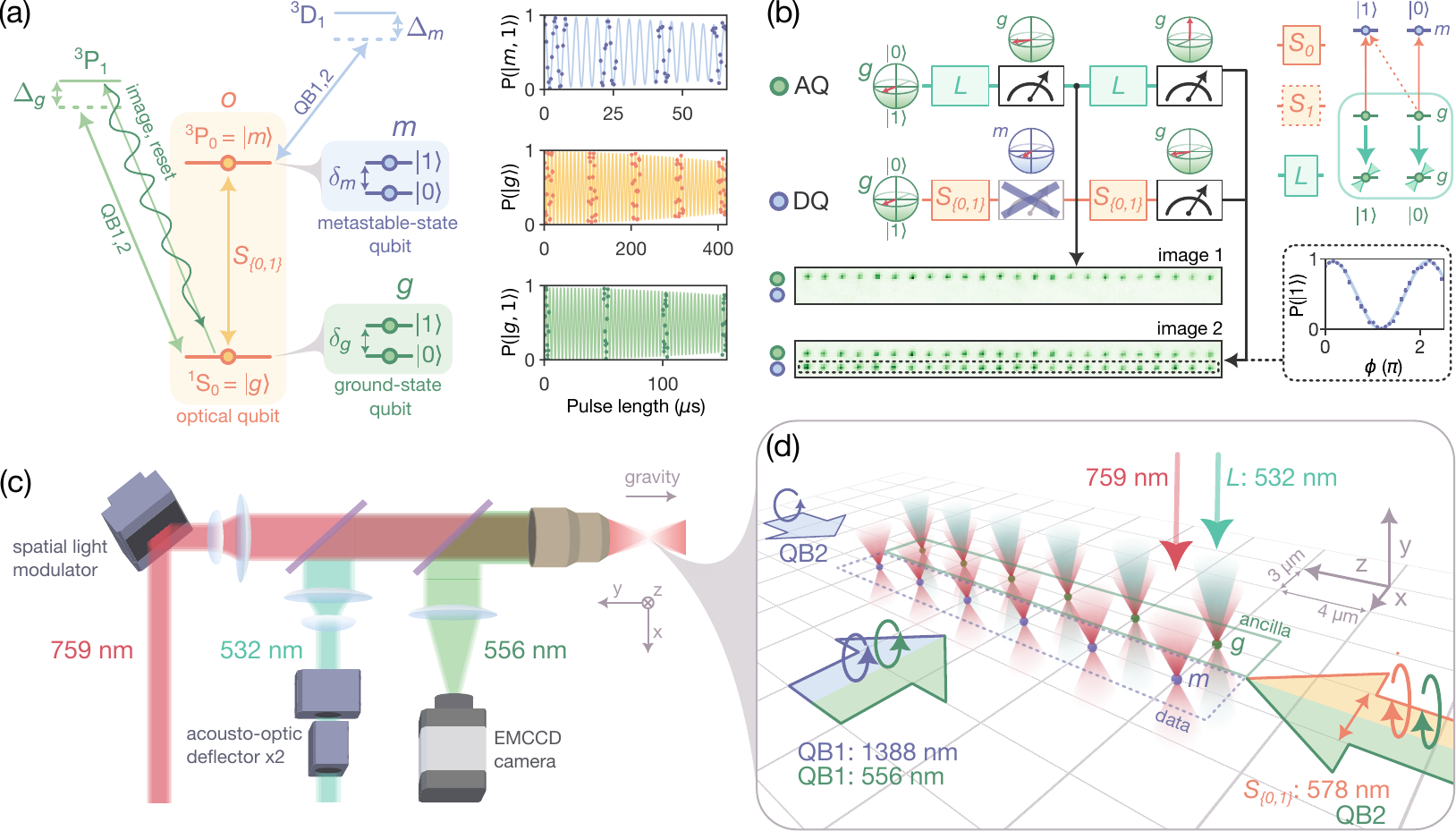}
\caption{\label{fig:apparatus}Mid-circuit operations using the \textit{omg}-architecture in ${}^{171}$Yb. (a) The \textit{omg} level structure in ${}^{171}$Yb: the optical qubit $o$ is defined on the clock transition between${}^1\mathrm{S}_0$ and ${}^3\mathrm{P}_0$ states; the metastable $m$ and ground $g$ state qubits are defined by the $\up \equiv |m_F=+1/2\rangle$ and $\down \equiv |m_F=-1/2\rangle$ Zeeman states of nuclear spin-1/2 within the respective manifolds. The optical Raman transitions through ${}^3\mathrm{P}_1$ and ${}^3\mathrm{D}_1$ drive nuclear qubit rotations in $g$ and $m$. The insets show Rabi oscillations with $\{\Omega_{m}/2\pi, \Omega_{o}/2\pi, \Omega_{g}/2\pi\}  \approx \{\RabiM, \RabiClockSigma,  \RabiG\}$ and $1/e$ damping of $\approx \{\flopsM, \flopsO, \flopsG\}$ cycles respectively. (b) Local shelving operations and MCO. The array is divided into data qubit (DQ) and ancilla qubit (AQ) subsets. The light shift operation $L$ hinders shelving for ancilla qubits leaving them in $g$ and therefore in a bright state for an image. The data qubits are shelved into $m$ with the $S_0$ shelving operation---a rotation on the $\pi$-polarized clock transitions which addresses both $\up$ and $\down$ states. At the end of MCO, the data qubit is assessed for coherence (right inset). The complimentary $S_1$ operation---a rotation on the $\sigma^-$-polarized clock transition---shelves only $\up$ spin state, realizing a non-destructive spin-sensitive detection. Atom images shown are averages of 600 shots. (c) Trapping, imaging and local addressing apparatus. (d) The qubit addressing beam geometry. QB1,2 control $R_X$ and $R_Z$ nuclear spin qubit rotations in ground the $g$ (green) and metastable $m$ (blue) manifolds. The $S_{0,1}$ beam controls optical qubit and shelving operations between the two nuclear spin qubit manifolds. The 532 nm tweezers ($L$) site-selectively light shift the clock transition for local shelving operations.  }
\end{centering}
\end{figure*}

The two latter approaches rely on the data qubits being stored in states dark to the measurement operation. This capability also arises in the optical-metastable-ground state qubit (\textit{omg}) architecture (fig.~\ref{fig:apparatus}(a)), where the quantum information is encoded in either of the three qubit manifolds: optical $o$, ground $g$ and metastable $m$. Within this architecture, quantum information can be coherently transferred between the manifolds, with certain (possibly mid-circuit) operations acting only on one of the manifolds. While the concept of \textit{omg} originally surfaced for trapped ions systems~\cite{allcock2021omg,yang2022realizing,bazavan2023synthesizing,debry2023experimental}, tweezer-trapped alkaline-earth-like atoms such as ytterbium-171 (${}^{171}$Yb) are also ideal candidates for this architecture~\cite{jenkins2022ytterbium, ma2021Universal, chen2022analyzing}. For applications in quantum information science, the \textit{omg} architecture pairs well with already demonstrated features of ${}^{171}$Yb, including efficient preparation of low-entropy arrays, fast, high-fidelity single qubit gates, non-destructive state detection, as well as Rydberg-mediated two-qubit gates~\cite{gorshkov2009alkaline, jenkins2022ytterbium, ma2021Universal,huie2023repetitive, ma2023high}.  Indeed, in very recent related work, detected leakage from the metastable states was used to improve post-selected performance~\cite{scholl2023erasure}, as well as facilitate mid-circuit erasure conversion~\cite{ma2023high}. 

In this work, we demonstrate full control of the \textit{omg} architecture in ${}^{171}$Yb and exploit its features to realize MCOs, including both measurements and resets, in a tweezer array of 48 sites. Here, the two Zeeman levels in each of the $g$ and $m$ manifolds serve as the nuclear spin qubit basis with $\up \equiv |m_F=+1/2\rangle$ and $\down \equiv |m_F=-1/2\rangle$, such that the states available to the atom are $|\{{g,m}\},\{0,1\}\rangle$ (fig.~\ref{fig:apparatus}(a)). Within $g$ and $m$, we perform single qubit rotations on the nuclear spin qubit via optical Raman transitions, using beams globally illuminating the atom array, with single-qubit gate fidelity of $\mathcal{F}_{g} = \GroundQubitFidelity{}$ and $\mathcal{F}_{m} = \MetastableQubitFidelity{}$. In both manifolds, we observe coherence times of many seconds, due to an absence of coupling between the nuclear and electronic spin~\cite{jenkins2022ytterbium, ma2021Universal, ma2023high}. Further, we demonstrate global and local control of the optical qubit with single-qubit fidelity of $\mathcal{F}_{\o} = \OpticalQubitFidelity{}$ for global operations. We exploit the optical qubit for rapid, high-fidelity shelving operations between the $g$ and $m$ manifolds, and describe a new method that minimizes recoil heating from the shelving. With rotations on the $\sigma^-$-clock transition, we turn $|{g},0\rangle$ dark to the imaging light and reinitialize it afterwards in $g$. We distinguish the two spin states with detection fidelities of $\mathcal{P}_{|{g,0\rangle}} = \UpFidelity{}$ and $\mathcal{P}_{|{g,1\rangle}} = \DownFidelity{}$, successfully reinitializing 97.4(3)\% and 99.0(2)\% of $\up$ and $\down$ atoms in the process, leaving them at low temperatures for further operations. 

With the addition of light-shifting tweezers, we realize high-fidelity atom-selective shelving that enables MCOs. In protocols of approximately $20$ ms duration, we demonstrate mid-circuit state-readout of the ancilla qubits, and, in a separate experiment, reset of the ancillas' spin and motional degrees of freedom via Raman-sideband cooling. For ground-state qubits, these operations preserve the state of data qubits at the $95\%$ level, with a realistic path to $>99\%$ performance. Emphasizing the versatility of this approach, we further show how these tools can be used to implement mid-circuit measurement on the metastable qubit.  

\section{\label{sec:apparatus}Experimental apparatus}

To prepare arrays of ${}^{171}\mathrm{Yb}$, we use an apparatus previously described in~\cite{jenkins2022ytterbium}.  For MCO, our basic strategy is to realize site-resolved shelving operations of data qubits into the metastable space, and then perform operations on the ancilla qubits in the ground-state space by applying  near-resonant light to the $^3\mathrm{P}_1$ states (fig.~\ref{fig:apparatus}(b)). Accordingly, successful protection of the data qubits relies on (1) the fidelity of the local shelving operation, i.e. clock rotations, and, (2) that the time for ground state operations, such as atom detection, must be short compared to the decay time of qubit coherence in the metastable space. We implement several significant experimental upgrades to realize this scheme and satisfy these constraints.

First, we add an optical tweezer trapping potential at 759 nm (fig.~\ref{fig:apparatus}(c,d)). At this wavelength, the polarizabilities for $|{g,\{0,1\}}\rangle$ and $|{m,\{0,1\}}\rangle$ are equal (\enquote{magic} condition), eliminating tweezer-depth-dependent light shifts and thus improving control of the clock transition~\cite{ye2008quantum, ludlow2015optical}. The tweezer array geometries are generated by reflecting the 759 nm beam from a spatial-light-modulator (SLM), which imprints a programmable phase pattern onto the light. By subsequently focusing the beam through a 0.6 NA microscope objective, the phase information is translated into an intensity pattern of our choice ($2 \times 24$ tweezer array for these experiments)~\cite{Nogrette2014single,barredo2016atom}. 

Secondly, we implement a fast, high-fidelity and low-loss scheme for imaging on ${}^1\mathrm{S}_0\leftrightarrow{}^3\mathrm{P}_1$ in 759 nm tweezers. For the transition to $|{}^3\mathrm{P}_1 ,~F'=3/2,~ m_{F'}=-1/2\rangle$, we find a \magicangle angle between the quantization axis and the tweezer light polarization (along $x$), where the differential polarizability of the ground and excited states vanishes~\cite{norcia2018microscopic}. This magic condition decouples scattering and cooling rates from the atom position within the trap, minimizing loss during imaging. To image and cool, we address the atoms with two beams red-detuned from the resonance, with a net $k$-vector that has projection along all trapping axes. With that, in 3.5 ms we collect $\approx$20 photons per atom and focus them onto an Electron Multiplying Charge-Coupled Device (EMCCD) camera (fig.~\ref{fig:apparatus}(c)). We distinguish an atom and a vacancy with \imgfidelity{} fidelity, while losing \imgloss ~of atoms per image, predominantly due to Raman scattering of 759 nm photons to ${}^3\mathrm{P}_2$ -- a state anti-trapped at 759 nm. 

Preparation of motional ground state aids the fidelity of operations on the clock transition. As such, we realize improved gray-molasses cooling (GMC) and Raman-sideband cooling in 759 nm tweezers, initializing the atom in $\bar{n}\approx 0.05$ motional state within 4 \ms. Additionally, the combination of GMC and blue-shielding \cite{brown2019gray,jenkins2022ytterbium}, allows us to load single atoms into 759 nm tweezers with up to \loadeff{} efficiency. 

With the ability to perform fast detection and ground-state cooling in magic-wavelength tweezers, we can apply clock light at 578 nm to drive rotations on the optical qubit (fig.~\ref{fig:apparatus}(d)). We operate in the regime where the clock Rabi frequency $\Omega_{o}/(2\pi) \approx \RabiClockSigma,$ is much larger than the radial trapping frequency $\omega/(2\pi) = \ramseytrapfreq$~\cite{campbell2010ultrafast}. Here, the population transfer between $g$ and $m$ occurs with high fidelity, reduced sensitivity to atomic motional state and at speeds comparable to those of our single qubit gates~\cite{jenkins2022ytterbium,chen2022analyzing}. Depending on the experiment, we use circular or linear polarizations (with the quantization axis along $z$ or $x$) that respectively drive a $\sigma^-$ transitions ($|g,0\rangle\leftrightarrow |m,1\rangle$) or a $\pi$ transitions $|g,i\rangle\leftrightarrow |m,i\rangle$ for $i\in\{0,1\}$). For pulse lengths of $\pi / \Omega_{o}$, these rotations constitute shelving operations denoted here as $S_1$ and $S_0$ respectively (fig.~\ref{fig:apparatus}(b)). During $S_0$, the quantization axis is defined by a 1.5 G magnetic field, which splits the nuclear spin qubits by $\delta_g=\QBsplittingG$ and $\delta_m=\QBsplittingM$. For $S_1$, we increase the field to 32 G, to provide additional frequency selectivity for the $\sigma^-$ transition. For shelving operations, we operate with $\Omega_{o}/(2\pi) \approx 80 \kHz$.

To achieve atom-resolved control of the optical qubit, we employ 532 nm tweezers generated with acousto-optic-deflectors (AODs) (fig.~\ref{fig:apparatus}(c,d)). These tweezers are used for local light-shifting operations $L$, that take the clock transition out-of-resonance for selected sites (fig.~\ref{fig:apparatus}(b)). The resultant site-selective shelving between $g$ and $m$ allows for single-qubit gates, measurement and reset operations to gain local character. 

\begin{figure*}
\begin{centering}
\includegraphics{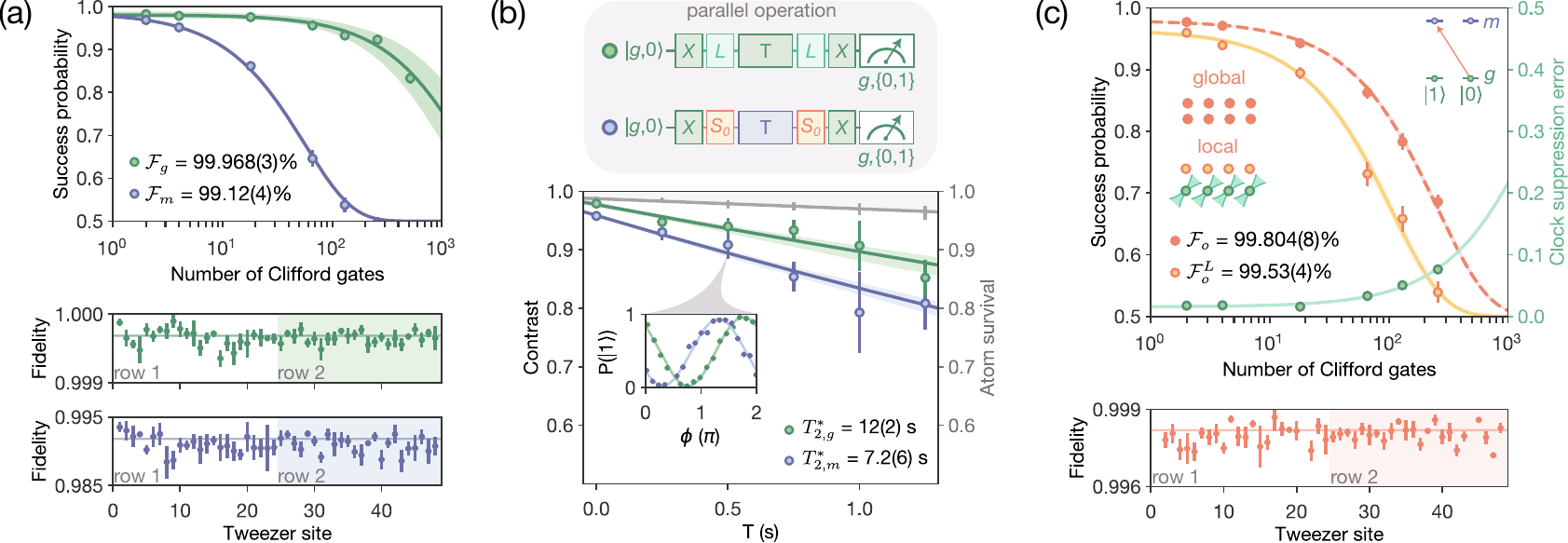}
\caption{\label{fig:control} Characterization of \textit{omg} qubits and their control. (a) Clifford randomized benchmarking (RB) for $g$ (green) and $m$ (blue) qubit manifolds. The Clifford gates are compiled from $X\equiv R_X(\pi/2)$ and $Z\equiv R_Z(\pi/2)$ rotations, with typical pulse lengths $\gatelength$. The magnetic field during the nuclear spin qubit RB is 1.6 G. We measure average single qubit gate fidelities of $\mathcal{F}_{g} = \GroundQubitFidelity$ and $\mathcal{F}_{m} = \MetastableQubitFidelity$. (bottom) Site-wise analysis of average single-qubit fidelities. (b) Coherence measurement for $\g$ (green) and $\m$ (blue). In a Ramsey-type experiment, contrast is recorded for dark times in $g$ and $m$, while scanning the phase $\phi$ accumulated between $\up$ and $\down$ qubit states. We extract $T_{2,g}^* = \groundcohdecay$ and $T_{2,m}^* = \metacohdecay$ respectively. (c) RB for optical qubit $o$, performed without light shifting beam (global) and in the presence of light shifting beam (local). In the former case, we measure average single qubit gate fidelities of $\mathcal{F}_{o} = \OpticalQubitFidelity$, which decreases to $\mathcal{F}_{o}^L = \OpticalQubitFidelityLS$ in the presence of the light shifting beam. (green) The extracted error in clock suppression is $\ClockSuppressionError$ per Clifford gate. (bottom) Site-wise analysis of average single-qubit fidelities (three tweezer sites not included due to insufficient statistics).
}
\end{centering}
\end{figure*}

Lastly, we add the capability to perform global single-qubit operations in both $g$ and $m$ nuclear spin manifolds. Following the scheme outlined in our prior work~\cite{jenkins2022ytterbium}, we realize an $R_X$ nuclear qubit rotation by driving an optical Raman transition via an intermediate state (${}^3\mathrm{P}_1$ for $g$ and ${}^3\mathrm{D}_1$ for $m$). Figure~\ref{fig:apparatus}(d) shows the beam geometry. Here, a single beam of circular polarization propagating perpendicular to the quantization axis provides all necessary components of the Raman transition while benefiting from motional-state insensitivity and the intrinsic phase stability. To drive an $R_Z$ qubit rotation, we employ a beam of the same wavelength and polarization but propagating along the quantization axis, which induces a light shift between the qubit states. Due to the symmetry between the two beams (QB1,2), we can control whether a beam performs $R_X$ or $R_Z$ rotation with the direction of the quantization axis. To minimize Raman scattering from the qubit beams, we operate at large detunings with $\Delta_{g}/(2\pi) = \DeltaG$ and $\Delta_{m}/(2\pi) = \DeltaM$ (from ${}^3\mathrm{P}_1, F'=1/2$ and ${}^3\mathrm{D}_1, F'=3/2$ respectively).

\section{\label{sec:omg}Control of the \emph{OMG} architecture}

The nuclear qubit is initialized in the $|{g},0\rangle$ state, by optically pumping on the $|g,1\rangle \leftrightarrow |{}^3\mathrm{P}_1,~ F'=1/2,~ m_{F'}=+1/2\rangle$ transition \cite{jenkins2022ytterbium}. A subsequent shelving operation can then change the qubit manifold to $m$ if desired. We detect the nuclear spin state following two approaches: (1) ejecting atoms in $|{g},0\rangle$ from the traps through rapid driving on the stretched transition $|{g},0\rangle\leftrightarrow  |{}^3\mathrm{P}_1,~F'=3/2,~ m_{F'}=+3/2\rangle$ \cite{jenkins2022ytterbium}; or (2) hiding $|g,0\rangle$ from (or exposing $|m,1\rangle$ to) imaging light through $S_1$ shelving operation. The results for the non-destructive approach are presented in sec.~\ref{sec:nddetect}.

Figure~\ref{fig:control}(a) presents the characterization of the average single-qubit gate fidelity for $g$ and $m$ with Clifford randomized benchmarking (RB)~\cite{Knill2008Randomized}. The Clifford gates are compiled from native gates, $X\equiv R_X(\pi/2)$ and $Z\equiv R_Z(\pi/2)$, through a software package \texttt{pyGSTio}~\cite{pygsti}, with $\approx$3.5 native gates per Clifford gate~\cite{pygsti, jenkins2022ytterbium, ma2021Universal}. The gate sequences of varying lengths are then applied to atoms and their measured final state is compared to the ideal outcome; for each circuit depth, forty random Clifford strings are generated, with ideal outcome randomized between $|0\rangle$ and $|1\rangle$. Single qubit fidelities are extracted from fits to the success probability as a function of circuit depth, revealing $\mathcal{F}_{g} = \GroundQubitFidelity$ and $\mathcal{F}_{m} = \MetastableQubitFidelity$. We also analyze RB data site-wise, revealing homogeneous fidelities across the array, the spread being consistent with statistical variation. The error sources are discussed in Appendix~\ref{sec:rberrors}. 

Next, we compare the nuclear qubit coherence in the $g$ and $m$ manifolds in a Ramsey-type experiment (fig.~\ref{fig:control}(b)). With an $X$-gate acting on $|{g},0\rangle$, we prepare a $\down_y = \frac{1}{\sqrt{2}}(\up -i\down)$ superposition in $g$, and subsequently apply $S_0$ to half of the atoms, shelving those qubits into $m$. After variable dark time $T$, we bring the qubits back to $g$ and complete the Ramsey sequence for both qubit subsets with another $X$-gate. We scan the phase $\phi$ accumulated between $\up$ and $\down$ and record the contrast of the resultant fringe. The decay at short times of the extracted coherences are consistent with $T_{2,g}^* = \groundcohdecay$ and $T_{2,m}^* = \metacohdecay$, where the timescale for the metastable qubit is limited by additional decay mechanisms. 

Finally, we characterize the control of the optical qubit and shelving operations. The optical qubit is initialized in $|g,\{0,1\}\rangle$ and the state-readout is performed by measuring the $|g,\{0,1\}\rangle$ population, while the $|m,\{0,1\}\rangle$ population is dark to the imaging light (${}^1\mathrm{S}_0\leftrightarrow{}^3\mathrm{P}_1$). A major detection error here is the dark-to-bright state leakage due to Raman scattering of 759 $\nm$ photons:  ${}^3\mathrm{P}_0$$\rightarrow$${}^3\mathrm{S}_1$$\rightarrow$${}^3\mathrm{P}_1$, which then decays to ${}^1\mathrm{S}_0$~\cite{dorscher2018lattice}. While at the trap-depths employed for qubit operations ($U/h = \ramseydepth$), this scattering rate is $\ramseyscattrate$, for the trap-depths employed in imaging ($U/h = \imgdepth$), that rate increases to $\imgscattrate$. If the atom decays to the ground state and scatters enough collected photons to cross the detection-threshold, an error will occur. 

Similar to the nuclear spin qubit, we perform RB and extract $\mathcal{F}_{\o} = \OpticalQubitFidelity$ average fidelity of single-qubit operations in the $\o$ manifold (fig.~\ref{fig:control}(c)). Here $X$ and $Z$ gates, realized with $\pi/2$ clock rotations on the $\sigma^-$-transition and $90\degree$ laser phase jumps respectively, are used to compile the Clifford gates, with $\approx 1.7$ $X$ gates per Clifford gate. The observed error rate is consistent with the calculated error per Clifford gate present in $\Omega_{\o}\gg \omega$ regime (see Appendix~\ref{sec:rberrors}). This error arises from off-resonantly driving multiple motional sidebands \cite{poyatos1996trapped, campbell2010ultrafast} and reduces with increasing the $\Omega_{\o}/ \omega$ ratio \cite{jenkins2022ytterbium}. We also note that $\pi$-polarization impurity present in the clock drive would manifest as a non-Markovian error, which is not appropriate for RB characterization. We estimate this effect to contribute errors an order of magnitude smaller than the motional errors. 

We characterize the performance of local optical qubit rotations by performing RB for $\o$ in the presence of light-shifting tweezers applied to half of the array. For the non-light-shifted subset of sites, we observe an optical qubit fidelity of $\mathcal{F}_{o}^L = \OpticalQubitFidelityLS$ per Clifford gate, the additional error likely arising due to finite overlap with diffracted or scattered 532 nm light; similar level of performance for local optical qubit operations was observed in recent work \cite{shaw2023multi}. We also analyze the suppression of clock rotations for the light-shifted sites, extracting $\ClockSuppressionError$ suppression error per Clifford gate applied to the other subset of sites. Since we operate in the $\Omega_{\o}\gg \omega$ regime, our measured coherence timescales of the optical qubit are on the scale of $2\pi/\omega$, due to coupling of the motional- and spin-states of the atom.  In addition, the local clock operations are constrained to cases when the qubits on light-shifted sites are either in $g$ or $m$, since otherwise the light-shift operation would destroy any coherences. 

\begin{figure}
\begin{centering}
\includegraphics{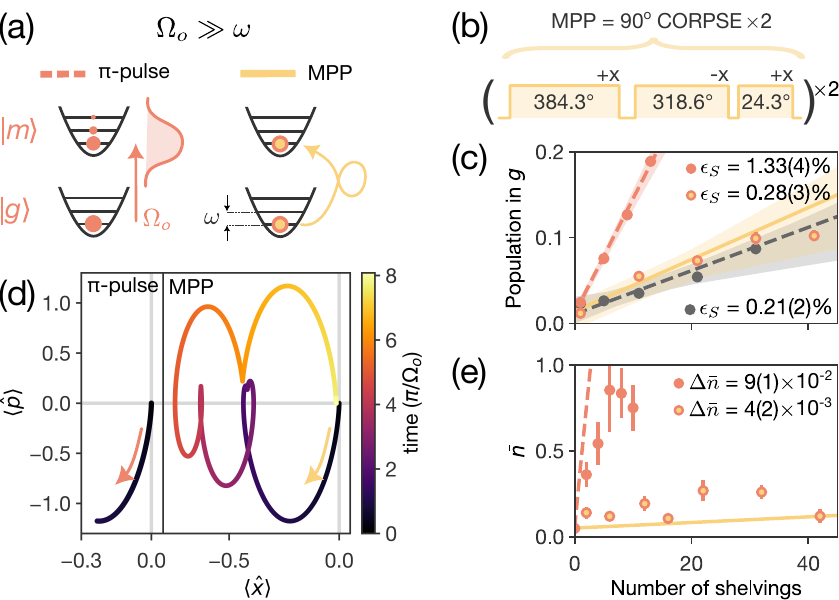}
\caption{\label{fig:shelve} Coherent shelving. (a) Motional effects in the $\Omega_{\o}\gg\omega$ regime: a single $\pi$-pulse addresses multiple sidebands heating up the atom, while MPP leaves the original motional state unchanged after the operation is complete. (b) Pulse sequences for the MPP. MPP comprises of two CORPSEs, each realizing a target $90\degree$ rotation on the clock transition. One CORPSE consists of $384.3\degree_{+x}$, $318.6\degree_{-x}$ and $24.3\degree_{+x}$ rotations, with $\pm x$ denoting $0\degree$ and $180\degree$ phases of the applied pulses \cite{cummins2003tackling}. (c) Population remaining in the ground state after odd number of shelving operations. From the slope we extract the error per shelving operation: $\epsilon_S = 0.21(2)\%$ for a $\pi$-pulse and no wait time (grey); $\epsilon_S = 1.33(4)\%$ for a $\pi$-pulse and 2 ms wait time (orange); and $\epsilon_S = 0.28(3)\%$ for MPP and 2 ms wait time. (d) Simulation of the motional state evolution in the $\langle\hat{x}\rangle-\langle\hat{p}\rangle$ phase space of a harmonic oscillator under a $\pi$-pulse (left) and MPP (right) with the units of a zero-point motion.  (e) Average number of motional quanta vs. number of shelving operations. A $\pi$-pulse and MPP increase $\bar{n}$ by $9(1)\times10^{-2}$ and $4(2)\times10^{-3}$, respectively. Lines show a no-free-parameter simulation of the expected behaviour for the two cases (see Appendix~\ref{subsec:MPPsim}).
}
\end{centering}
\end{figure}

For MCOs, we employ the above methods for rapid shelving operations between the ground and metastable manifolds. An imperfect shelving operation will leave a small atom population ($\epsilon_S$) in the original manifold. The major source of this error is the already mentioned motional effect: a clock $\pi$-pulse excites a superposition in the motional degree of freedom while off-resonantly driving motional sidebands (fig.~\ref{fig:shelve}(a)). Since these are coherent effects, for a single tweezer, one can always find a $2n\pi/\omega$ wait time between two shelvings (for integer $n$) for which $\epsilon_S$ is minimized (ignoring trap anharmonicity). However, the distribution of trapping frequencies among the tweezers in the array makes the above approach unfeasible. For wait times of the order of few ms, the coherent effects average-out leading to $\epsilon_S>1\%$ (fig.~\ref{fig:shelve}(c)). We reduce this error 5-fold with a motional-state preserving pulse (MPP), a composite pulse sequence which controls the motional degree of freedom during the shelving operation. The MPP consists of two consecutive CORPSE (Compensation for Off-Resonance with a Pulse SEquence) pulses \cite{cummins2003tackling}, each realizing a target $90\degree$ rotation on the clock transition (fig.~\ref{fig:shelve}(b)). The origin the improvement in the shelving performance for the MPP becomes clear while looking at the evolution in the $\langle\hat{x}\rangle-\langle\hat{p}\rangle$ phase space of the trap harmonic oscillator (fig.~\ref{fig:shelve}(d)). For a $\pi$-pulse the motional state is coherently displaced away from the origin, while MPP traces a closed loop, with the state arriving back at the origin at the end of the pulse sequence. We note that related ideas are used in trapped-ion two-qubit gates to remove spin-motional entanglement~\cite{choi2014optimal}. We assess the MPP performance by looking at the average number of motional quanta $\bar{n}$, accessible to us through sideband spectroscopy, as a function of the number of shelvings applied (fig.~\ref{fig:shelve}(e)). We find that the atoms are heated 20-times less with MPP compared to a single $\pi$-pulse.  

\section{\label{sec:nddetect}Clock-based non-destructive state-readout of $^{171}$Yb}

The \textit{omg} architecture in ${}^{171}$Yb naturally enables non-destructive state read-out, in which the state of the nuclear spin qubit is determined in a manner that does not depend on destruction of one spin component~\cite{barnes2021assembly}, as is often done~\cite{wilk2010entanglement,isenhower2010demonstration,kaufman2012cooling,levine_parallel_2019,jenkins2022ytterbium,ma2021Universal}. The ability to reuse the measured qubits can significantly improve experimental duty-cycle, which is key for high precision experiments, and can also improve the resilience of atomic clocks against local oscillator noise~\cite{dick1989local, schioppo2017ultrastable, norcia_seconds-scale_2019, schulte2019prospects}. Similarly, in quantum error-correction protocols, it is desirable to salvage measured ancilla qubits so that they may be recycled for subsequent code cycles. The potential benefits of non-destructive state detection have motivated multiple demonstrations in neutral atom arrays~\cite{gibbons2011nondestructive, fuhrmanek2011free,kwon2017parallel,martinez2017fast,shea2020submillisecond, chow2022high}, including nuclear spin qubits of $^{87}$Sr and very recently $^{171}$Yb~\cite{barnes2021assembly,huie2023repetitive}. In this work, using high-fidelity shelving operations, we demonstrate clock-based non-destructive nuclear-qubit detection with low loss, and we identify clear steps to improve the scheme to $>$99\%. Importantly, compared to alkali demonstrations employing repeated scattering on cycling transitions, our methods leave the temperature of the measured qubits $<$$10$ $\upmu$K, so that they may be readily used for subsequent operations.

\begin{figure}
\begin{centering}
\includegraphics{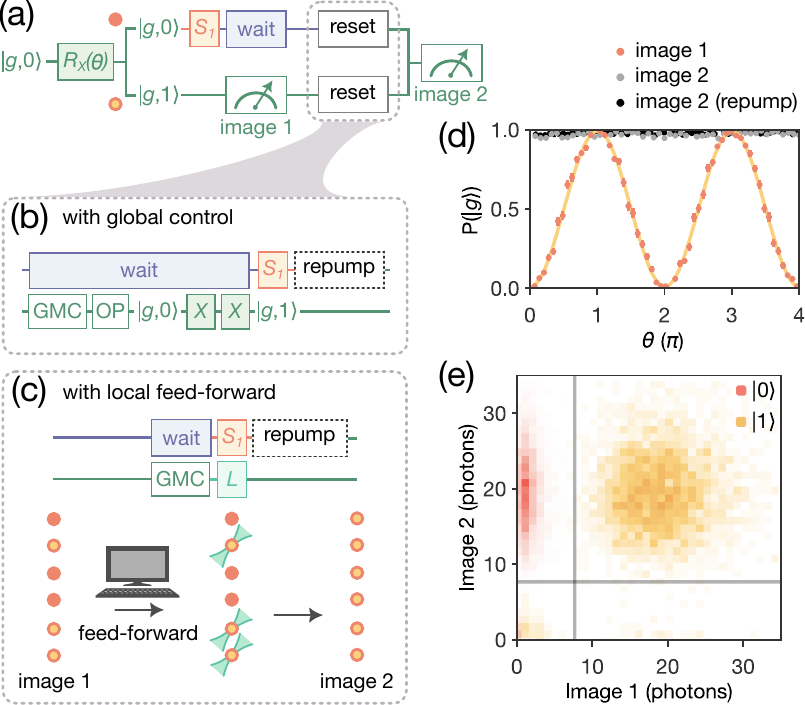}
\caption{\label{fig:nddetect}Non-destructive state-sensitive detection. (a) Overview of the protocol. $S_1$ turns spin $\vert g, 0 \rangle$ dark, shelving it to the $m$ manifold, while $\vert g, 1 \rangle$ remains and appears bright for the image. The spins are then reset back to the $g$ manifold. A final image checks for atom loss at the end of the procedure. (b) and (c) show the two reset approaches employed. (d) Image 1 (yellow) is spin sensitive, resolving $\up$ and $\down$ populations of a $|\psi\rangle = \cos{\frac{\theta}{2}} \up -i \sin{\frac{\theta}{2}}\down$ input state in $g$. Image 2 shows the atoms can be reinitialized in $g$ following spin detection via global control. No correction was applied to the data shown. (e) 2D histogram showing counts recorded in both images for atoms initially prepared in $\up$ (orange) and $\down$ (yellow). Reinitialization in $g$, for this data set, was performed via global control and repumping step before image 2.}
\end{centering}
\end{figure}

Figure~\ref{fig:nddetect}(a) shows the outline of the procedure. We start by preparing a qubit in the $g$ manifold in $|\psi\rangle = \cos{\frac{\theta}{2}} \up -i \sin{\frac{\theta}{2}}\down$ state. We then apply an $S_1$ shelving operation and take an image. Since $|g,0\rangle$ population is shelved into $|m,1\rangle$, it appears dark, while the $|g,1\rangle$ population appears bright. Subsequently, we perform a reset, bringing the shelved population back to $g$, and take a second image to assess whether the atoms were reinitialized in the ground state at the end of the procedure. Figure~\ref{fig:nddetect}(d) shows typical measurement outcomes for the two images. 

We explore two schemes for performing reset: (1) a sequence of global operations and (2) a local feed-forward approach~\cite{huie2023repetitive,singh2023mid}, which each have benefits depending on experimental aims. In the former (fig.~\ref{fig:nddetect}(b)), the atoms measured in the bright state are reinitialized in $|g,1\rangle$ through optical pumping followed by a $R_X(\pi)$ rotation. Alternatively, in the second reset scheme (fig.~\ref{fig:nddetect}(c)), we use processed information from the first image to apply a local light-shift onto sites that were detected bright. In both cases, a subsequent $S_1$ operation addresses only the atoms that appeared dark in the first image, resetting all into $g$. The two reset schemes suffer from errors that may leave atoms in $m$ at the end of the sequence. As such, to distinguish these errors from atom loss from traps, we include an additional repumping step through ${}^3\mathrm{D}_1$ before the second image.    

To assess detection fidelities of correctly identifying $\up$ as dark and $\down$ as bright, we independently prepare and measure each spin state (fig.~\ref{fig:nddetect}(e)), and extract $\mathcal{P}_{|g,0\rangle} = \UpFidelity{}$ and $\mathcal{P}_{|g,1\rangle} = \DownFidelity{}$ while post-selecting on atom survival at the the end of the sequence. The analysis method and infidelity contributions are presented in detail in Appendix~\ref{subsec:ndsd} and Table~\ref{tab:ndsd}. For the $\up$ state, the major error sources include Raman scattering of 759 nm photons in deep traps during imaging and shelving error, while the $\mathcal{P}_{|g,1\rangle}$ is mainly limited by polarization impurity of the clock drive. 

We also compare the probabilities for successfully resetting the atoms to $g$ for the two schemes employed (see Appendix \ref{subsec:ndsd} and Table~\ref{tab:ndsd}). The global approach, with atoms repumped at the end of the sequence, performs the best, preserving $p_{|g,0\rangle} = \UpNDprob{}$ and $p_{|g,1\rangle} = \DownNDprob{}$ of the two populations. The losses here are dominated by vacuum lifetime and Raman scattering to anti-trapped ${}^3\mathrm{P}_2$. With future technical improvements, we project that reset probabilities of $\{99.1\%, 99.9\%\}$ for $\{\up, \down\}$ are within reach, which would be on par with the state-of-the-art~\cite{chow2022high}. For some applications, repumping may not be desireable (e.g. if the data qubit is being stored in $m$ at the same time). We find that without repump the feed-forward approach is better at resetting $|g,1\rangle$ population back to $g$. However, due to errors stemming from detection infidelity in image 1 and Raman scattering during image-processing time, this method resets a smaller fraction of $|g,0\rangle$ atoms at the same time. %Nevertheless, the ability to apply feed-forward is a necessary ingredient for error-correction schemes, and we hope to expand this capability further in the future.

\begin{figure*}
\begin{centering}
\includegraphics{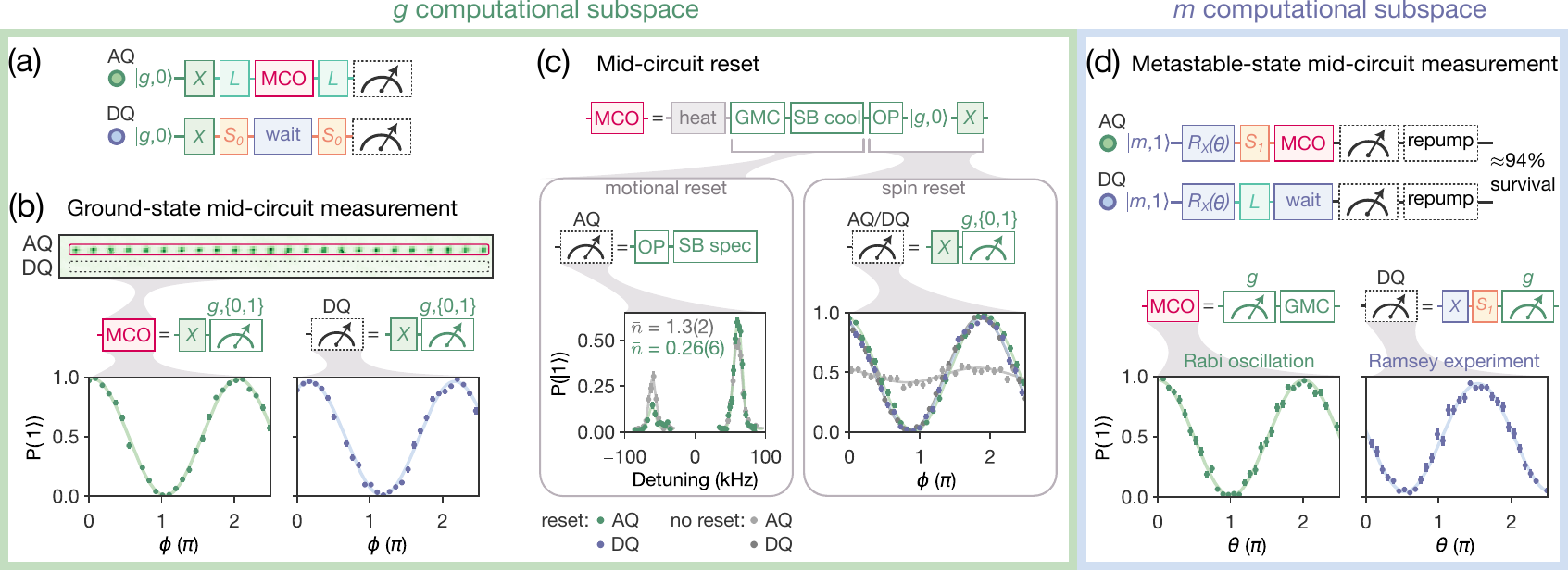}
\caption{\label{fig:mco} Mid-circuit operations (MCOs). (a) Overview of the protocol. With local light-shifts $L$ acting on ancilla qubits (AQ), $S_0$ shelves data qubits (DQ) to the metastable state, dark to MCOs. After the MCO is completed, the DQ is brought back to the ground state and assessed for coherence loss. For reset operations, a final measurement is performed on AQ to check whether the operation was successful. (b) Mid-circuit measurement results. In a Ramsey-type experiment, the contrast for AQ is measured mid-circuit with $\mathcal{C}_\mathrm{MCM}^\mathrm{AQ} = \MCMAQcontrast$. After the MCO, the contrast for the DQ is $\mathcal{C}_\mathrm{MCM}^\mathrm{DQ} = \MCMDQcontrast$. Atom image shown is an average of 600 shots. (c) Mid-circuit reset results. Prior to reset we apply a pulse of a 399 nm heating beam, which increases $\bar{n}$ and mixes the spin state. The gray-molasses cooling (GMC) and Raman sideband cooling (SB cool) then reset the motional state, while optical pumping (OP) and an $X$-gate, reset the spin state of AQ. (left) Sideband spectroscopy (SB spec), showing decrease in $\bar{n}$ after reset. (right) Ramsey-type experiment, showing contrast reset for AQ to $\mathcal{C}_\mathrm{{MCR}}^\mathrm{AQ} = \MCRAQcontrast$. The DQ contrast after the MCO was $\mathcal{C}_\mathrm{{MCR}}^\mathrm{DQ} = \MCRDQcontrast$. (d) Starting from qubits stored in the $m$ manifold, a Rabi oscillation of a $\mathcal{C}_{\mathrm{MCM},m}^\mathrm{AQ} = 96.2(8)\%$ contrast and a Ramsey experiment with a $\mathcal{C}_{\mathrm{MCM},m}^\mathrm{DQ} = 90(1)\%$ contrast are performed for AQ and DQ, respectively. The state-readout for AQ and DQ is non-destructive, retaining $\approx$94\% of atoms at the end of the sequence. For all panels, no corrections are applied to the data nor to the reported contrasts; see text for an explanation of SPAM and errors.
}
\end{centering}
\end{figure*}

\section{\label{sec:mco}Mid-circuit operations}

Equipped with the capabilities offered by the \textit{omg} architecture, we realize mid-circuit operations (MCOs) on this platform. We divide our system into two subsets of ancilla (AQ) and data qubits (DQ), and perform operations on the ancilla qubits, while preserving the coherences stored in the data qubits. Figure~\ref{fig:mco}(a) presents the schematic diagram of our procedure. We start by initializing all atoms in a $\down_y = \frac{1}{\sqrt{2}}(\up -i\down)$ state within the $g$ manifold and follow by a $S_0$ shelving operation on the data qubits to the $m$ manifold. Subsequently, we perform MCOs on ancillas, including state-sensitive measurement and reset of the spin and motional degrees of freedom, and then unshelve the data qubit. In the reported results, the mid-circuit operations take approximately $20$ ms, which reflects the time between shelving and unshelving of the data qubit. We characterize the coherence of the data qubits following the MCOs via a Ramsey-type protocol. Depending on the experiment, we also take a final measurement of ancilla qubits to confirm if the implemented MCO was successful. 

In the first experiment, we perform mid-circuit measurements (MCMs) on ancilla qubits (fig.~\ref{fig:mco}(b)). In a Ramsey-type experiment, we scan the phase $\phi$ accumulated between $\up$ and $\down$ and measure a contrast of $\mathcal{C}_\mathrm{MCM}^\mathrm{AQ} = \MCMAQcontrast$. The observed contrast reduction is consistent with known errors (see Table \ref{tab:mco}). The state-dependent readout employed here is destructive to the $|g,0\rangle$ state; this temporary limitation can be relieved with more laser power and separate beams for $S_0$ and $S_1$ operations. After the mid-circuit measurement, we follow with a contrast measurement for the data qubit obtaining a raw $\mathcal{C}_\mathrm{MCM}^\mathrm{DQ} = \MCMDQcontrast$, which is uncorrected for $2.0(2)\%$ state-preparation and measurement (SPAM) errors. The main contributors to the data qubit contrast reduction are: Raman scattering in deep tweezers during imaging and shelving errors (Table \ref{tab:mco}). 

As a separate method to characterize the performance of the MCM, we quantify dephasing, depolarization, and rotation errors through quantum process tomography \cite{chuang1997prescription, jevzek2003quantum} on both the data and ancilla qubits (Appendix \ref{subsec:QPT}). We reconstruct the MCM process, excluding loss and shelving errors, and find process fidelities $F_{p,\mathrm{DQ}} = 0.972(5)$ and $F_{p,\mathrm{AQ}} = 0.979(6)$ for the data and ancilla, respectively. These process fidelities can be converted into average state fidelities $F_{av}$ that reflect loss and shelving errors \cite{bhandari2016general}, for which we find $F_{av,\mathrm{DQ}} = 0.961(3)$ and $F_{av,\mathrm{AQ}} = 0.972(4)$, consistent with the Ramsey measurements. 

The second experiment, mid-circuit reset, consists of two parts: reset of the motional state and reinitialization into the desired nuclear spin qubit state (fig.~\ref{fig:mco}(c)). Our demonstration starts with a heating beam that drives the 399 nm ${}^1\mathrm{S}_0\leftrightarrow {}^1\mathrm{P}_1$ transition. A 150 ns pulse is enough to controllably increase the average number of motional quanta to $\bar{n}=\MCOnoresetnbar$, while also creating a mixed state in the spin degree-of-freedom. We then reset the $\bar{n}$ with a combination of gray-molasses cooling and Raman sideband cooling. A subsequent optical pumping and an $X$-gate prepare the initial $\down_y$ state of the ancilla.  We note that the time required to perform gray-molasses cooling, sideband cooling and optical pumping (all in deep tweezers) is 3.8 ms -- comparable to the imaging time. After the reset is completed, we perform two separate measurements on the ancilla: Raman sideband spectroscopy to measure $\bar{n}$ and the ancilla qubit coherence measurement through a Ramsey sequence. With the reset, we observe a reduction in ancilla's temperature to $\bar{n} = \MCOresetnbar$ and increase in the fringe contrast from $\MCOnoresetAQContrast$ to $\mathcal{C}_\mathrm{reset}^\mathrm{AQ} = \MCRAQcontrast$. Finally, we record the contrast for the data qubit, following the reset operation, obtaining $\mathcal{C}_\mathrm{reset}^\mathrm{DQ} = \MCRDQcontrast$. The contrast reduction sources are listed in Table \ref{tab:mco}. In addition to the reset operation, we note that the data qubit coherences were also not affected by the heating beam, pointing to the possibility of loading the broad-line magneto-optical trap in the presence of metastable spin qubits for atom replenishing. 

The MCO approach described so far assumed $g$ as the computational subspace for the data and ancilla qubits, with the $m$ manifold used exclusively for temporary data qubit shelving. This scheme is favorable for applications where the quantum information is to be stored for longer times, since it mitigates Raman scattering errors inherent to the $m$ manifold. However, a complimentary approach of using $m$ as the computational subspace may be preferred for certain tasks, and its use underlies recently proposed schemes for erasure conversion to improve logical qubit performance~\cite{wu2022erasure, ma2023high}. 

We employ the site-selective shelving techniques to realize MCOs also in the $m$ qubit manifold (fig.~\ref{fig:mco}(d)), which incorporates elements of the non-destructive state-readout discussed in sec.~\ref{sec:nddetect}. We start by preparing $|\psi\rangle = \cos{\frac{\theta}{2}} \up -i \sin{\frac{\theta}{2}}\down$ superposition with Raman rotations on the metastable qubit. An $S_1$ shelving operation on the ancilla qubit brings $|m,1\rangle$ population down to $g$ for measurement, while a local light shift $L$ acting on the data qubit shields it from shelving and therefore projection. We scan $\theta$ and record Rabi oscillation for the ancilla qubit with $\mathcal{C}_{\mathrm{MCM},m}^\mathrm{AQ} = 96.2(8)\%$ contrast. The MCO is concluded with recooling of the imaged ancilla atoms. To check whether coherence of the data qubit was preserved, we apply an $X$ gate to qubits in $m$ and non-destructively measure $|m,1\rangle$ population, completing a Ramsey-type experiment for the data qubits. The contrast of the Ramsey fringe, recorded by scanning $\theta$, is $\mathcal{C}_{\mathrm{MCM},m}^\mathrm{DQ} = 90(1)\%$. Finally, we repump all atoms to $g$ and measure $\approx$94\% survival of both ancilla and data atoms. The breakdown of the errors for this protocol is presented in Table~\ref{tab:mco}, with the Raman scattering being the dominant error source.

\section{\label{sec:discussion}Discussion}

Moving forward, we can identify several areas in which to improve and expand the MCO performance.  Errors from Raman scattering can be nearly eliminated with differential control of the tweezer depths, so that data qubits are kept in shallow traps. Our total shelving error exceeds the theoretical prediction for the motional-state preserving pulse ($\approx 10^{-4}$) substantially, which will be the subject of future investigation (see Appendix~\ref{subsec:MPP}). To scale to 100s of qubits, we will need roughly 10 times more power with similar phase noise properties, which is likely achievable with newly available fiber-laser products. Further, with modest beam path changes, the reported mid-circuit measurement of $g$ qubits can be made non-destructive using a combination of $S_1$ and $S_0$ operations. Accordingly, with realistic upgrades, we expect then that this versatile shelving approach to MCO can reach above $>99\%$ efficacy as well as perform non-destructively for both $g$ and $m$ qubits. %As an additional benefit, these capabilities of high-fidelity, site-selective shelving of large qubit arrays will also readily enable \emph{local} two-qubit gates from the metastable state for atoms' within a blockade radius, without the need for atom moves~\cite{bluvstein2022quantum}. 

As an alternative to clock rotations, both the shelving and Raman scattering errors might be avoided by leaving data qubits in the ground-state. In this case, it would be possible to use a combination of $^3\mathrm{P}_1$ light-shifting (i.e. with 680 nm on the $^3\mathrm{S}_1\leftrightarrow {}^3\mathrm{P}_1$ transition) and non-destructive spin-detection in the ground-state~\cite{huie2023repetitive, AC2023} to perform site-selective measurements. This comes at the price of more stringent fractional field stability, as such methods rely on large magnetic fields~\cite{huie2023repetitive,AC2023}. Perhaps, most importantly, using a large field splits the nuclear qubit in such a manner that it is more challenging to operate in a regime where the Rabi-frequency significantly exceeds the qubit frequency~\cite{jenkins2022ytterbium}. Consequently, one must operate in the opposite, more typical regime with slower single-qubit gate operations. In addition, the motional reset via Raman sideband cooling is likely incompatible with  $^3\mathrm{P}_1$ light shifting, since the Raman transition would not be suppressed in the same manner as resonant driving of the $^3\mathrm{P}_1$ transition. Looking forward, using a combination of $^3\mathrm{P}_1$ and $^3\mathrm{P}_0$ light-shifting might be a particularly powerful suite for MCO. 

The realization of high-fidelity mid-circuit operations in a neutral atom array, and, in particular $^{171}$Yb, has immediate implications. Neutral atom arrays possess many favorable features for quantum-information processing: single-site addressibility and readout, real-time adjustment qubit connectivity, and tunable range Rydberg interactions~\cite{xia2015randomized,saffman2016quantum, graham2022multi, bluvstein2022quantum}.  The recent realization of the two-qubit entangling gates with fidelities of $\lukinFidelity$ in alkali qubits, and 98\% in $^{171}$Yb nuclear-spin qubit, bring neutral atom arrays even closer to implementing fault-tolerant computation~\cite{evered2023high, ma2023high}. When combined with high fidelity two-qubit gates, the parallel, high-fidelity MCOs demonstrated here pave the way for first realizations of quantum error correction in a neutral atom array.

Mid-circuit operations can also enable protocols that enhance metrology. For Ramsey interferometry in atomic clocks, it is known that mid-circuit measurement can be used to track phase slips and thereby extend the interrogation time for which a Ramsey signal can be inverted for laser frequency correction~\cite{rosenband2013exponential,Borregaard2013Efficient, zheng2023reducing, shaw2023multi}. In this context, the use of sub-ensembles with cascaded interrogation times, potentially incorporating mid-circuit measurement and/or site-resolved clock rotations, can allow an improvement in atom-laser stability that is exponential in the number of sub-ensembles~\cite{rosenband2013exponential,Borregaard2013Efficient,KohlhaasPhase2015,shaw2023multi,zheng2023reducing}. To achieve Heisenberg-limited metrology in the presence of laser phase noise, mid-circuit measurement can be deployed for phase tracking of rapidly evolving Greenberger–Horne–Zeilinger (GHZ) states, while the addition of feedback can be used to preserve the enhanced signal-to-noise of squeezed states~\cite{andre2004stability,borregaard2013near, kessler2014heisenberg,schulte2019prospects,pezze2020heisenberg}. By shelving optical qubit coherence in the nuclear qubit via the tools demonstrated here, these protocols become realizable in an atom array optical clock~\cite{norcia_seconds-scale_2019,madjarov_atomic-array_2019,young_half_2020}. In addition, the use of mid-circuit reset could allow for near continuous operation of an optical atomic clock, in a manner that mirrors multi-apparatus clock experiments~\cite{schioppo2017ultrastable}. 

Mid-circuit measurement and reset also increasingly play a role in many-body physics. A new class of phase transitions  have garnered recent significant interest, where a transition in entanglement scaling is determined by the relative rate of mid-circuit measurements compared to coherent dynamics~\cite{skinner2019measurement, potter2022entanglement, fisher2023random}. Combining the tools demonstrated in this work with Rydberg-mediated interactions are one potential path to realizing these models experimentally~\cite{balewski2014rydberg, zeiher2017coherent, eckner2023realizing, bluvstein2022quantum, graham2022multi}. In a similar vein, the use of mid-circuit reset might open new paths to dissipative preparation of long-range entangled states or even autonomous error correction, in an atom array of 100s of qubits~\cite{tantivasadakarn2021long,iqbal2023topological,reiter2017dissipative}.

Beyond applications enabled by MCO, the MPP-based shelving protocol we report provides a rapid method for \emph{repeatedly} and \emph{locally} transferring quantum amplitudes between the $g$ and $m$ manifolds, which has important implications for using the $omg$ architecture for neutral-atom quantum computing. In particular, the single-photon Rydberg transition from the metastable manifold has been the basis of recent demonstrations of high fidelity Rydberg physics and two-qubit gates in alkaline-earth atoms~\cite{madjarov2020high, schine2021long, eckner2023realizing,scholl2023erasure,ma2023high}. Paired with atom-selective MPP shelving, the two-qubit gates and Rydberg-mediated operations can be locally controlled, and used repeatedly in deep quantum circuits. Meanwhile, untargeted qubits can remain in the $g$ manifold, which is well-isolated from the environment and easily controlled. These methods therefore make it possible to fully exploit the \emph{omg} architecture, and the flexibility it affords for optimizing different state manifolds for distinct quantum operations. 

\section{\label{sec:conclusions}Conclusion}

In this work, we have demonstrated control of the \textit{omg} architecture in neutral ${}^{171}$Yb atoms trapped in optical tweezer arrays. We characterized the single-qubit manipulation of the ground, metastable and optical qubits, and compared the coherence timescales of the ground and metastable nuclear spin qubits. We have shown non-destructive state-detection, with atom reset aided by local feedback, and mid-circuit measurement and reset, enabled by local shelving operations between the ground and metastable states. We expect that these tools will have manifold applications in quantum science, as well as underlie new technical capabilities, such as site-resolved two-qubit gates. 

\textit{Note}: During completion of this work, we became aware of related studies in $^{171}$Yb from Atom Computing~\cite{AC2023}.

\section{\label{sec:acknowledgements} Acknowledgements}

We wish to acknowledge Hannes Bernien, Nelson Darkwah Oppong, Matteo Marinelli, and Ana Maria Rey for a careful reading of this manuscript. We also thank Felix Veitmayer for assistance with FPGA integration in the control system, and Andrew Ludlow for helpful conversations regarding our clock laser system.  These results are based upon work supported by the Office of Naval Research (N00014-20-1-2692), Air force Office of Scientific Research (FA9550-19-1-0275), Army Research Office (W911NF-19-1-0149, W911NF-19-1-0223), National Science Foundation Physics Frontier Center (Phys-1734006), U.S. Department of Energy, Office of Science, National Quantum Information Science Research Centers, Quantum Systems Accelerator, and the National Institute of Standards and Technology. This research was supported by an appointment to the Intelligence Community Postdoctoral Research Fellowship Program at JILA at the University of Colorado, Boulder administered by Oak Ridge Institute for Science and Education (ORISE) through an interagency agreement between the U.S. Department of Energy and the Office of the Director of National Intelligence (ODNI). A. S. acknowledges the support from the Funai Overseas Scholarship. W. M. acknowledges support of the NIST NRC program. F. R. acknowledges the support from Cluster of Excellence Matter and Light for Quantum Computing (ML4Q) EXC 2004/1–390534769.

\appendix

\renewcommand{\thefigure}{A\arabic{figure}}
\setcounter{figure}{0}

\section{\label{sec:experiment_sequence}Method}

\subsection{\label{subsec:experiment_sequence}Overview of the experimental procedure}
In this work, $^{171}\mathrm{Yb}$ atoms are trapped in a $2\times24$ tweezer array generated by a spatial light modulator (SLM). The trapping light is at 759 nm, the magic wavelength for the ytterbium clock transition, ${^1}\mathrm{S}_0\leftrightarrow{}{^3}\mathrm{P}_0$. Atoms are loaded from a narrowline magneto-optical trap (MOT) at 556 nm (${^1}\mathrm{S}_0\leftrightarrow{}{^3}\mathrm{P}_1$) into the tweezers, initially loading several atoms per tweezer. Employing the same transition, we realize light-assisted collisions and leave each tweezer with either 0- or 1-atom occupancy. We achieve enhanced loading through the same scheme as described in our previous work~\cite{jenkins2022ytterbium}, though this time we load into the tweezers at 759 nm instead of 532 nm. We find that the loading scheme achieves a lower loading fraction for similar densities, up to \loadeff{}, and demands longer loading times. 

After loading the tweezer array, an initial image is taken to identify which tweezers are occupied. With gray molasses cooling followed by Raman sideband cooling, we reduce the atom temperature, initializing the atoms near to the radial motional ground state~\cite{jenkins2022ytterbium}. (see sec.~\ref{subsec:cooling} for details). Finally, the atoms are prepared in the $\ket{g, 0}$ state, with a fidelity of 99.72(5)\%, by optically pumping on the $\ket{{^1}\mathrm{S}_0,~m_F=-1/2}\leftrightarrow\ket{{^3}\mathrm{P}_1,~F'=1/2,~ m_{F'}=+1/2}$ transition. Destructive spin detection in the $g$ manifold is realized by detuning the optical pumping beam to resonantly drive the $\ket{{^1}\mathrm{S}_0,~m_F=+1/2}\leftrightarrow\ket{{^3}\mathrm{P}_1,~F'=3/2,~ m_{F'}=+3/2}$ transition, thereby heating atoms in $\ket{g, 0}$ out of the traps. The fidelity of this process is $\PushOutFidelity$, with the residual error stemming from a small fraction of $|g,1\rangle$-state atoms being pumped to $|g,0\rangle$ by the push-out beam. 

To minimize Raman scattering of trap light from the clock state, the main experiment is typically conducted in 230 kHz (11~$\upmu$K) shallow tweezers, with only imaging and cooling performed in 8.7 MHz (0.4 mK) deep tweezers. In the experiments, images are taken to determine which tweezer sites have an atom in the $^1\mathrm{S}_0$ state following atom manipulations. See sec.~\ref{sec:timing_diagram} for further details regarding the experimental sequence. 

\subsection{\label{sec:MagicAngle}Magic angle and fast imaging}
\begin{figure}
\centering
\includegraphics{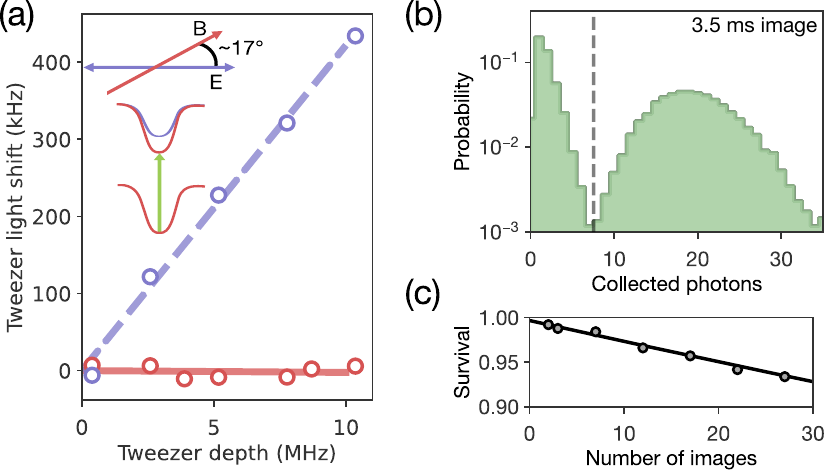}
\caption{\label{fig:magicangle} Magic angle and fast imaging. (a) Magic angle condition for the ${^1}\mathrm{S}_0\leftrightarrow|{^3}\mathrm{P}_1,~F'=3/2,~m_{F'}=-1/2\rangle$ transition. Purple points are measured with the magnetic field (B) parallel to the tweezer polarization (E), and red points, with B tilted by 17$\degree$ with respect to E (near magic angle condition for our configuration). The tweezer-induced light shift has a linear sensitivity to the trap intensity that is about 50 times smaller for the magic angle condition than for the parallel case. (b) Histogram for high-fidelity fast imaging. The imaging has a duration of 3.5 ms and a discrimination fidelity of 99.80(5)\%. Dashed line is the photon count threshold---events with photon counts below (above) the threshold are identified as dark (bright). (c) The survival rate after a variable number of images, yielding an average loss of 0.19(2)\% per image after accounting for the vacuum loss.}
\end{figure}

Although 759 nm is a magic wavelength for the clock transition, it is not for the ${^1}\mathrm{S}_0\leftrightarrow{}{^3}\mathrm{P}_1$ transition, which we use for imaging. To realize fast imaging of the ground state, we find a first-order magic condition for the $|^1\mathrm{S}_0\rangle\leftrightarrow|{^3}\mathrm{P}_1,~ F'=3/2,~ m_{F'}=-1/2\rangle$ transition (fig.~\ref{fig:magicangle}(a)). When the magnetic field ($B=16$ G) is tilted by $\approx$17$\degree$ with respect to the tweezer polarization, the perturbation from the tweezers mixes $F'=3/2$ states~\cite{norcia2018microscopic,saskin2019narrow}, such that the sensitivity of the transition frequency to trap intensity is reduced to much less than the transition's natural linewidth over the range of trap intensities employed.

In this condition, we perform fluorescence imaging using two non-retroreflected beams, propagating along the axial and the radial direction of the tweezer respectively. Both beams have an intensity of several tens of $I_\mathbf{sat}$ and are red-detuned by several $\Gamma$ from the transition. $I_\mathbf{sat} = 0.14 ~\mathrm{mW/cm^2}$ and $\Gamma/(2\pi)= 180~ \mathrm{kHz}$ is, respectively, the saturation intensity and natural decay rate of the $^1\mathrm{S}_0\leftrightarrow{^3}\mathrm{P}_1$ transition. We find that an imaging duration of 3.5 ms enables a discrimination fidelity of 99.80(5) \% (fig.~\ref{fig:magicangle}(b)). To assess the imaging loss, we take a variable number of consecutive images and record the final atom survival (fig.~\ref{fig:magicangle}(c)). After accounting for vacuum losses occurring during the wait times in-between the images ($\approx$20 ms), we extract a loss rate of 0.19(2)\% per image, which is consistent with the expected Raman scattering rate of atoms from $^3\mathrm{P}_1$ into the anti-trapped $^3\mathrm{P}_2$ state.

\subsection{\label{subsec:cooling}Gray molasses cooling and Raman sideband cooling}

\begin{figure}
\centering
\includegraphics{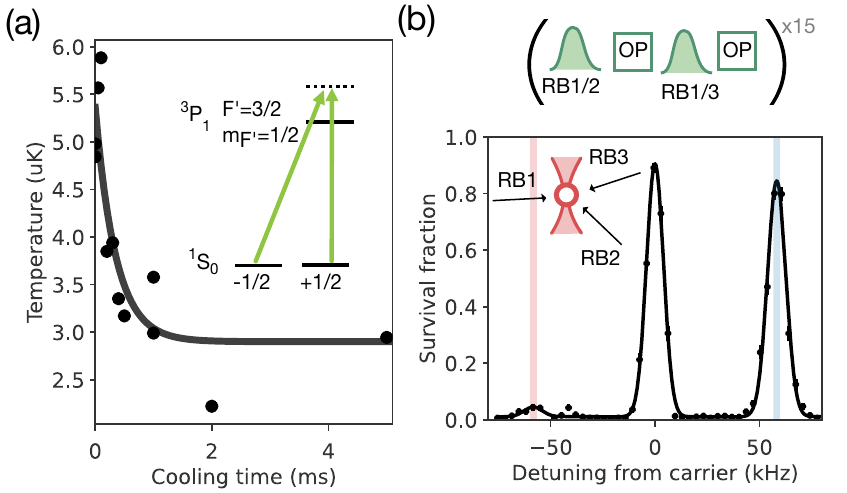}
\caption{\label{fig:method} Gray molasses cooling and Raman sideband cooling. (a) Temperature of tweezer-trapped atoms as a function of gray-molasses cooling time. The temperature is determined by comparing results of a release-and-recapture experiment and a Monte Carlo simulation for an atom of a certain temperature \cite{tuchendler2008energy}. The black line is an exponential fit, yielding a cooling time constant of 0.4(2) ms and a final temperature of 3 $\upmu$K. (inset) Magnetic field splits excited states of ${}^3\mathrm{P}_1$ realizing a lambda configuration employed for the gray-molasses cooling. (b) (top) Pulse sequence for Raman sideband cooling in the radial direction. The two axes of the radial direction are cooled alternately, each typically for 15 cooling cycles. One cooling cycle consists of a Gaussian-shaped Raman pulse and a rectangular optical pumping pulse. (bottom) Raman sideband spectroscopy for a probing time corresponding to a $\pi$-pulse on the sideband ($\approx 3\pi$-pulse on the carrier). The average motional state, $\bar{n}$, is estimated to be about 0.05 from the ratio of the blue and red sidebands.}
\end{figure}

We perform gray molasses cooling (GMC) with the same beams as the 556 nm 3D MOT~\cite{jenkins2022ytterbium}, and with the same magnetic field as used in imaging. With detuning of $\approx9\Gamma$ from the ${^1}\mathrm{S}_0\leftrightarrow$ $|{^3}\mathrm{P}_1,~F'=3/2,~m_{F'}=+1/2\rangle$ transition and a total beam intensity of 76 $I_\mathrm{sat}$, we reach $3~\upmu$K temperatures in less than 1 ms, as shown through release-and-recapture measurements in fig.~\ref{fig:method}(a). Independently, through Raman sideband spectroscopy, we measure an average motional quanta after GMC to be $\bar{n}\approx 0.5$ for the trap frequency of 58 kHz. This corresponds to a temperature of 2.8 $\upmu $K---a value consistent with the release-and-recapture measurement. 

After gray molasses cooling, we use Raman-sideband cooling to further reduce temperature of the atoms in the radial dimension. We implement multiple cycles of pulsed Raman sideband cooling~\cite{jenkins2022ytterbium}. We operate with a trap frequency of $\omega_r/(2\pi) = 58$ kHz, a carrier Raman Rabi frequency of $\Omega_c/(2\pi) = 26$ kHz and Gaussian pulse shaping to reduce off-resonant excitation to other motional states. The pulse sequence and a Raman sideband spectroscopy following cooling are presented in fig.~\ref{fig:method}(b). For 15 cooling cycles (total cooling time of $\approx$2 ms), the atoms are prepared with an average motional quanta of $\bar{n}\approx 0.05$ in the radial direction. We note that even with only 5 cycles, the atoms can be cooled down to $\bar{n}\approx0.1$. This condition was preferred for the motional state reset experiment to shorten the total cooling time.

\section{\label{sec:Clock}Controlling the clock transition}

\subsection{\label{subsec:ClockLaser}Clock laser system}
\begin{figure}
\centering
\includegraphics[width=\columnwidth]{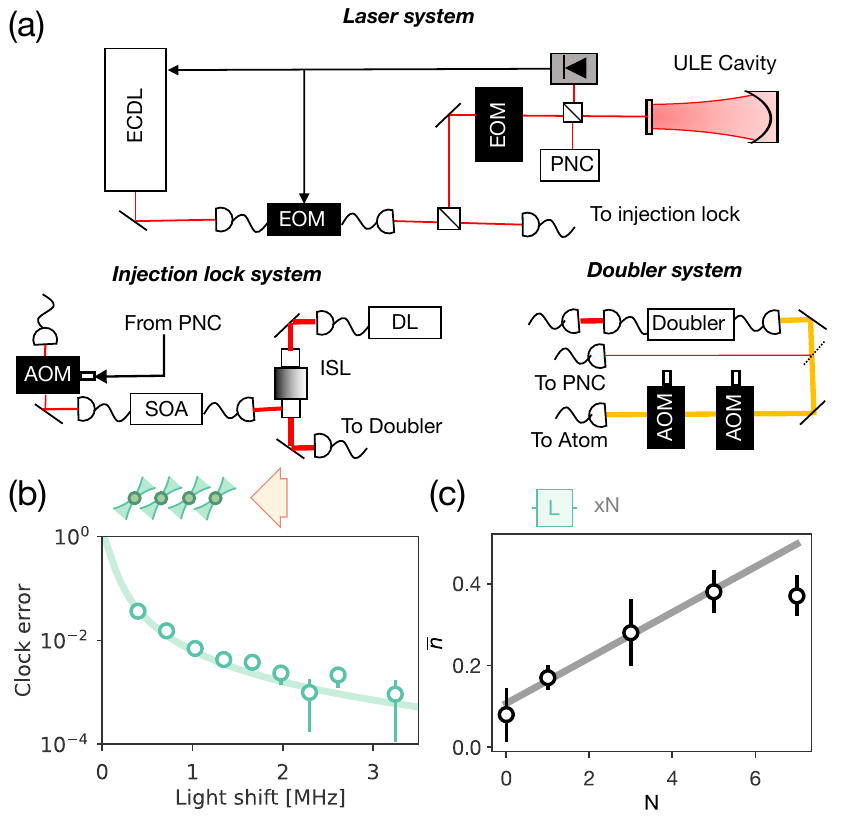}
\caption{\label{fig:clock} Control of the clock transition (a) Schematic diagram of the clock-transition laser system. 1156 nm seed laser light emitted from an external cavity diode laser (ECDL), locked to a high finesse ultra-low expansion (ULE) glass cavity, is pre-amplified by a semiconductor optical amplifier (SOA) and further amplified by an injection-lock to a high power DL in a butterfly package (DL), using a free-space optical isolator (ISL). Electro-optic modulators (EOMs) and acousto-optical modulators (AOMs) are mainly used for modulating frequency, except for the last AOM used for intensity servo. The frequency doubler is a single-path waveguide. Phase noise between the seed laser and the 578 distribution is removed with phase noise cancellation (PNC). (b) Suppression of excitation on the clock transition using 532 nm light. The ordinate is the probability that atoms are excited into $^3\mathrm{P}_0$ despite the presence of the 532 nm light, and the abscissa is the magnitude of the light shift from the 532 nm beam. A CORPSE-based motional state-preserving pulse (MPP) is applied here. The solid line is the Lorentzian line shape for the corresponding Rabi frequency, with no free parameters. (c) The heating by the light shift beam. The light shift operation $L$ is applied repetitively, and the final average motional quanta is measured via Raman sideband spectroscopy. From the linear fit (solid line) we extract 0.056(5) quanta accumulated per iteration of $L$. }
\end{figure}

The overview of the clock laser system is shown in fig.~\ref{fig:clock}(a). We lock an 1156 nm external cavity diode laser (ECDL) to an ultra-low-expansion (ULE) cavity with a finesse of 240,000. In addition to the laser current and laser piezo in the ECDL, we feedback on an in-loop fiber eom~\cite{endo2018residual}. This technique suppresses frequency noise near the clock Rabi frequencies used in this work ($\approx 100$ kHz)---difficult to achieve using only piezo and current feedback. After pre-amplifying the seed laser with a semiconductor optical amplifier (SOA), we injection lock the laser to a butterfly-packaged diode laser reaching $\approx$400 mW\cite{shimasaki2019injection}. The amplified laser is doubled by a single-pass waveguide doubler to obtain $\approx$90 mW of 578 nm light. Phase noise cancellation (PNC), referencing the beat note of the transmitted residual 1156 nm laser after the doubler and the initial seed laser, is used to suppress phase noise induced by fibers and amplifiers between the seed and the doubler.

\subsection{\label{subsec:ClockDelivery}Clock light delivery to the atoms}

Following the PNC pick-off, the clock light is intensity-stabilized to a photodiode and transduced with an acousto-optic modulator (AOM). A second AOM is used for fast pulse-shaping of the clock light. The light is delivered to the experiment on a short fiber (1 meter) and focused onto the atoms to the waist radius of $\approx20~\upmu$m. A piezo-actuated mirror near a Fourier plane is used for precise alignment of the spot onto the atom array, along with a separate mirror to optimize the dive angle. A Glan-Taylor polarizer (GTP), on an adjustable angle mount, is used as the last optic before the vacuum cell window to set the clock polarization. To optimize for a pure $\pi$-polarization component, as is desired for ground-state MCO operations, fine-alignment is accomplished by optimizing the magnetic field angle with three-axis control to minimize driving of a $\sigma$ transition. 

For experiments in which we drive $\sigma^-$ transitions, we add a quarter-wave plate after the GTP. The magnetic field quantization axis is angle optimized based on clock spectroscopy to purify the $\sigma^-$transition. For the non-destructive state detection, errors from the clock coupling to the untargeted spin component are a key concern, which arise due to non-zero $\Omega_{\sigma^+}$; for the optical qubit randomized benchmarking, contaminant $\Omega_{\pi}$ is most problematic, as it removes population from the two-level system of $\vert g, 0\rangle \leftrightarrow \vert m,1\rangle$. After optimization, we can directly spectroscopically measure the ratios $\Omega_{\sigma^+}/\Omega_{\sigma^-} = 0.02$ and $\Omega_{\pi}/\Omega_{\sigma^-} = 0.01$. These measurements inform the error budgeting discussed in Appendices~\ref{subsec:ndsd} and~\ref{sec:rberrors}.

\subsection{\label{subsec:LS}Local clock control}

For local control of the clock, we use a 532 nm tweezer array generated by crossed acousto-optical deflectors (AODs). We highlight the flexibility of this approach, in that the same set of tweezers can also be used to achieve $>90
\%$ loading effciency~\cite{jenkins2022ytterbium}. The crossed-AOD tweezer rail allows for fast, dynamic changes of the local light shift potential, a functionality which is not afforded by the SLM system supplying the 759 nm trap potential. We observe that a 532 nm tweezer of $\aodpower$ (at the microscope input) induces a light shift of 3 MHz. Figure \ref{fig:clock}(b) shows the measured clock suppression error when light-shifting is used in tandem with the motional-state-preserving pulse (MPP), described in the next section. We note that although the MPP is robust to small detuning errors, in the far-off-resonant regime, the (undesired) excitation rate generally falls off like a Lorentzian.

We observe the light shifting operation slightly heats up ground-state atoms each time it is applied. Figure \ref{fig:clock}(c) shows the measured average motional quanta after variable number of light-shifting operations, for the light-shift beam depth of 3 MHz and the beam intensity ramped linearly for 1.5 ms. The extracted heating rate is 0.06 quanta per operation. In the future, this could be improved by pulse shaping of the light-shifting beam.

\section{\label{sec:fit}Error bars and model fitting}
In this work, the error bars indicate a 1$\sigma$ confidence interval. Unless otherwise noted, the models are fitted with weighted least-squares methods, where the squares of the residuals are weighted by the inverse variances of the corresponding data points. In fig.~\ref{fig:control} and \ref{fig:shelve}(c), the shaded regions correspond to 1$\sigma$ confidence intervals, evaluated with Monte-Carlo methods given fit parameters and their covariance matrices.

\section{\label{sec:rberrors}Error assessment for $g$, $m$, and $o$ randomized benchmarking}

\begin{figure}
\centering
\includegraphics[width=\columnwidth]{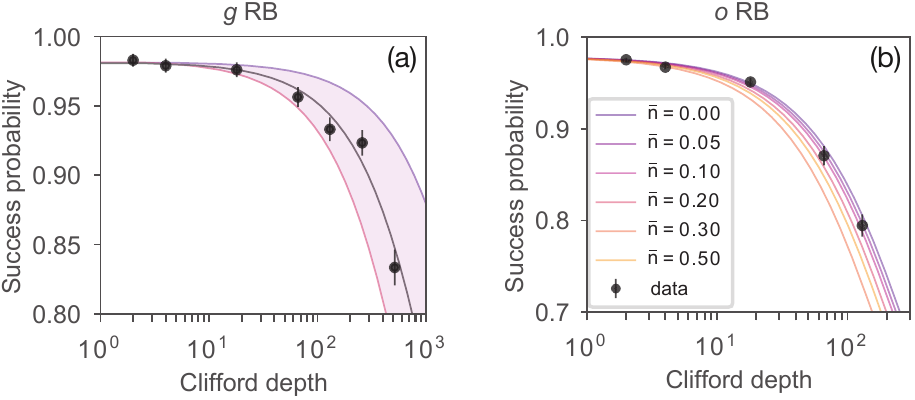}
\caption{\label{fig:RB_GS_clock} (a) Simulated randomized benchmarking in $g$ compared with measurements. We include measured intensity noise and site-to-site intensity variation, detuning errors, and scattering errors. We also include an error due to the $X$ and $Z$ beams propagating at an angle that is not perfectly orthogonal. Using the atoms as a probe, we measure the deviation from the orthogonal condition to be $\theta_X = 0.9(8)^\circ$. The total simulated error rate ranges from  $r=1.2\times10^{-4}$ at $\theta_X=0.1^\circ$ to $r=5.5\times10^{-4}$ at $\theta_X=1.7^\circ$ (shaded purple region). The black points are the measured values and the gray line is the fit to data. The success probability at 0 gates in the simulation is fixed to the value given by the fit to data. (b) Simulated randomized benchmarking of the optical qubits $o$ compared with measurements. Here we include only errors due to motional effects and plot the expected error as a function of initial $\bar{n}$. At the measured initial $\bar{n}=\nbar$, the calculated error rate $r=\OpticalQubitInFidelitySim$ due to motional effects is close to the measured value $r=\OpticalQubitInFidelity$. The black points are the measured values.}
\end{figure}

\textbf{Errors for the nuclear-spin qubit in \textit{g}.} A substantial source of error for the $g$ qubit arises due to the measured $0.8\%$ intensity variation from pulse-to-pulse fluctuations and $0.3\%$ intensity inhomogeneity across the array, together contributing an estimated error  of $\rbintflucerror$ per Clifford gate. The majority of the remaining error is likely due to the $\theta_X=0.9(8)^\circ$ measured deviation from orthogonality of our qubit beams. Raman and Rayleigh scattering from the intermediate states contribute an estimated error of $\rbgscatterr$. Other, smaller, sources of error include a detuning error due to the finite qubit splitting, which admixes a small $Z$-gate error into our $X$-gate. Figure \ref{fig:RB_GS_clock}(a) shows the simulated randomized benchmarking performance using the measured error rates and the same sets of gates used in the measured RB sequences.

\textbf{Errors for the optical qubit \textit{o}.}
Figure \ref{fig:RB_GS_clock}(b) similarly shows the simulated randomized benchmarking for the optical qubit. In this case, we find that the dominant error source comes from motional effects, and this is the only error simulated in this figure. We separately calculate the error rate due to polarization impurity and leakage to the states $|g,1\rangle$ and $|m,0\rangle$ and verify that this effect results in an error rate that is an order of magnitude smaller. In these simulations we use atoms consisting of the states $|g,0\rangle$ and $|m,1\rangle$, motional levels up to either $n=7$ or $n=11$, and we initialize the motional state at a temperature given by $\bar{n}$. We run the randomized benchmarking gate sequences for various starting temperatures and find that the motional errors at the measured initial value of $\bar{n}=\nbar$ are close to the total measured error rate in the experiment. We verify in the simulations that the probability of occupation of the largest simulated motional level ($n=7$, or $n=11$ for initial $\bar{n}>0.2$) remains under $5\times 10^{-3}$ for all gatesets. The final value of $\bar{n}$ also remains bounded for all gatesets to $\bar{n}<1.6$. 

\textbf{Errors for the nuclear-spin qubit in \textit{m}.} The error rate observed for $m$ is dominated by Raman scattering from the intermediate state. We directly measure population decaying to $\g$ while addressing $\m$ with the qubit beams during the randomized benchmarking sequence and extract the Raman scattering rate for ${}^3\mathrm{P}_0$$ \rightarrow$${}^3\mathrm{D}_1$$\rightarrow$${}^3\mathrm{P}_1$ process, obtaining $\Gamma_{{}^3\mathrm{P}_0 \rightarrow {}^3\mathrm{P}_1} = \MetaDecay $, which is near to a value calculated using known matrix elements. We compute the total decoherence rate from the scattering to be $\Gamma_{m,\mathrm{decoh.}} = \MetaDecoh $, with a corresponding error per Clifford gate of $\MetaError$, which is near to the measured value. In the future, we will improve $\mathcal{F}_{m}$ with higher optical powers and larger intermediate-state detunings. However, due to branching into other states of the ${^3}\mathrm{P}_J$ manifold, we anticipate that it will be challenging to achieve the error rates attained with the ground-state qubit.

\section{\label{subsec:MPP}Motional-state preserving pulse}

\subsection{\label{subsec:MPPtheory}Theoretical analysis}

To realize fast preparation of the clock state, we choose to operate in an unconventional regime where the clock Rabi frequency is larger than the tweezer trap frequency. A high clock-transition Rabi frequency is more acccessible for fermionic than bosonic alkaline-earth(-like) atoms, because of a relatively broader natural linewidth arising from the effect of hyperfine coupling~\cite{Xu2014Measurement}. Generally, the clock pulse modifies the atomic motional state even if the initial atomic state starts in the motional ground state, an effect which is exacerbated outside the resolved-sideband regime and for large Lamb-Dicke parameter. To address this, we find a condition in which such the motional state excitation is suppressed during the complete state transfer operation, which we call the motional-state preserving pulse (MPP).

\begin{figure}
\centering
\includegraphics[width=\columnwidth]{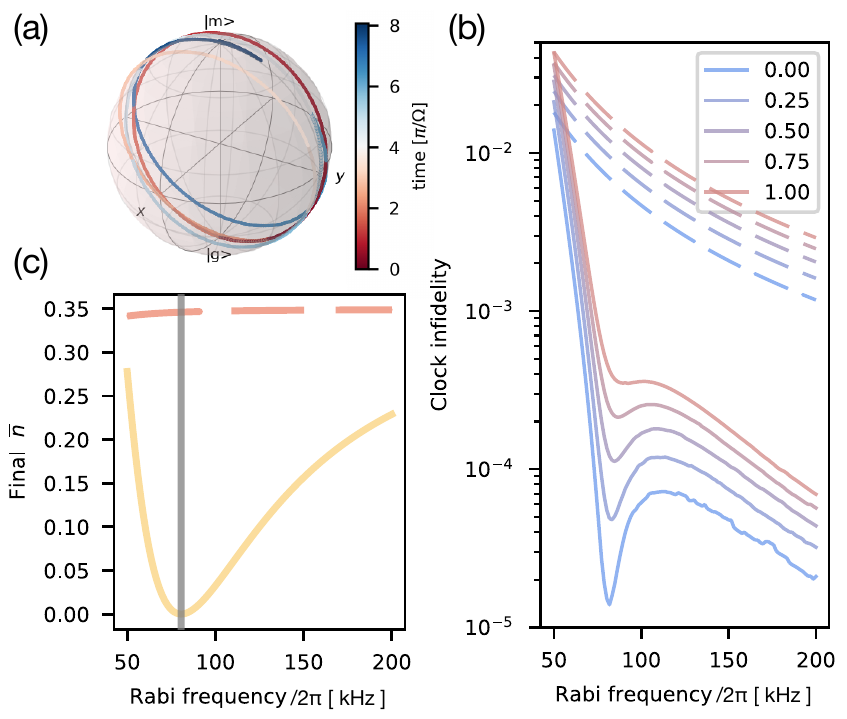}
\caption{\label{fig:MPP} Theoretical analysis of the motional state preserving pulse (MPP) in a harmonic oscillator. The trap frequency is fixed to be 10 kHz. (a) Bloch sphere picture of the state evolution in a single MPP. (b) Theoretical state transfer infidelity after a single MPP for various initial average motional quanta, indicating 2 orders of magnitude smaller infidelity, compared to a normal $\pi$-pulse. The dashed line is for the normal $\pi$-pulse and the solid line is for the MPP.  (c) Theoretical average motional state after a single MPP, assuming the initial state is in a motional ground state. A clear minimum can be observed at the Rabi frequency of $\approx$80 kHz, where we confirmed qualitatively this feature is retained in another choice of initial temperature. }
\end{figure}

We find that with a certain ratio of Rabi frequency to trap frequency, two $90$-degree CORPSE composite pulse sequences preserve the motional ground state while simultaneously realizing high fidelity state transfer. The Bloch sphere picture of state evolution with this pulse sequence is given in Figure~\ref{fig:MPP} (a), as well as the motional state evolution in Figure~\ref{fig:shelve} (d) in the main text. 

We analyze the MPP numerically for a trap frequency of 10 kHz, and the initial state is assumed to be the motional ground state. Fig. \ref{fig:MPP} (b) shows the theoretical estimation of the clock transfer fidelity for the MPP, which exceeds that of the normal $\pi$-pulse by nearly two orders of magnitude. Although the clock rotation fidelity has a flat dependence on Rabi frequency in this strong drive regime, the final average motional quanta with the MPP has a clear minimum at a specific Rabi frequency, as shown in Fig. \ref{fig:MPP} (c). We choose to operate all the shelving experiments near this optimal condition. Further exploration of the optimal pulse sequence for a shorter pulse, or a universal single qubit operation, is left for future work.

The observed MPP transfer fidelity in the experiment (0.28(3)\%) substantially exceeds the theoretical calculation ($\approx10^{-4}$). One possibility of the difference is the theoretical assumption of a perfect harmonic oscillator. We expect that anharmonicity of the Gaussian trap potential shape could influence the atomic motion in the shallow tweezer regime we have chosen to operate in. Errors from anharmonicity could be alleviated by using deeper traps while holding the ratio of Rabi frequency and trap frequency fixed. We leave this question to future investigations.

\subsection{\label{subsec:MPPsim}Simulation of motional heating}

In Figure~\ref{fig:shelve}(e) of the main text, we calculate the heating per shelving event for different shelving pulse sequences. When shelving atoms to the clock state, inhomogeneity of the trap frequencies across the array (typically with a fractional spread of 3.5\%) cause the motional states of different atoms to dephase rapidly compared to the timescale of the mid-circuit operations. In our clock shelving simulations, we account for this dephasing by removing motional coherences after each shelving step. After a clock pulse the atom states are given by $\rho$ with both motional and orbital degrees of freedom. In order to dephase the motional state, we take the partial trace over both degrees of freedom, $\rho_\mathrm{orbital}=\mathrm{Tr}_\mathrm{motional}(\rho)$, $\rho_\mathrm{motional}=\mathrm{Tr}_\mathrm{orbital}(\rho)$, and then keep only the diagonal components of the resulting motional state $\rho^D_{motional}$. Then we take the combined motional/orbital state to be the tensor product 
$$
\rho_\mathrm{orbital}\otimes\rho^D_\mathrm{motional}
$$

We simulate the accumulation of the average motional quanta by alternating this dephasing procedure and the unitary clock pulse evolution.

\section{\label{subsec:ndsd}Clock-based non-destructive state-readout} 

In the clock-based non-destructive state-detection experiment (sec.~\ref{sec:nddetect}) we take 3 images: image 0 (not included in fig.~\ref{fig:nddetect}) identifies sites with loaded single atoms; image 1 distinguishes between the two spin components $|g,0\rangle$ and $|g,1\rangle$; and image 2 assesses atom survival at the end of the sequence. In each image, the tweezer site appears either bright $b$ or dark $d$ and the measurement outcomes from the whole experiment are labelled $ijk$ with $i,j,k \in \{b,d\}$. In the following, we always post-select on $i=b$ (loaded tweezer sites). Since the atoms initialized in $|g,0\rangle$ and $|g,1\rangle$ are prone to different error and loss mechanisms, we extract detection fidelities and atom loss separately for each spin-state. In the experiment, we prepare atoms in state $s \in \{|g,0\rangle,|g,1\rangle\}$ and measure the number of $jk$ events for that state, $N_{jk}^s$.

\subsection{Detection fidelity}

The detection fidelities of correctly identifying $|g,0\rangle$ as $d$ and $|g,1\rangle$ as $b$, $\mathcal{P}_{|{g,0\rangle}}$ and $\mathcal{P}_{|{g,1\rangle}}$, are defined as
\begin{equation}
\mathcal{P}_{|{g,0\rangle}} = \frac{N_{db, c}^{|g,0\rangle}}{N_{db, c}^{|g,0\rangle}+N_{bb, c}^{|g,0\rangle}}
\quad\mathrm{and}\quad 
\mathcal{P}_{|{g,1\rangle}} = \frac{N_{bb, c}^{|g,1\rangle}}{N_{db, c}^{|g,1\rangle}+N_{bb, c}^{|g,1\rangle}}.
\end{equation}
Here we correct $N_{jk}^s$ for imperfect state-preparation via 
\begin{equation}
  \begin{pmatrix}
    N_{db, c}^s\\
    N_{bb, c}^s
  \end{pmatrix}
   =   
   \begin{pmatrix}
    1-\epsilon_\mathrm{OP} & \epsilon_\mathrm{OP}\\
    \epsilon_\mathrm{OP} & 1-\epsilon_\mathrm{OP}
  \end{pmatrix}^{-1}   
  \begin{pmatrix}
    N_{db}^s\\
    N_{bb}^s
  \end{pmatrix},  
\end{equation}
where $\epsilon_\mathrm{OP}$ is the optical pumping infidelity. Additionally, we post-select on atom survival in image 2 ($k=b$) to separate detection errors from atom loss. The analysis was conducted for experiments with reset via global control and a repump. The results are summarized in table~\ref{tab:ndsd}.

\subsection{Ground-state reset probability}

For the two reset approaches, (1) global control and (2) a local feed-forward, we calculate probabilities of re-initializing atoms back in $g$ following state-detection. For atoms prepared in $|g,0\rangle$ and $|g,1\rangle$ these are
\begin{equation}
p_{|{g,0\rangle}} = \frac{\sum_{j}N_{jb,c'}^{|g,0\rangle}}{\sum_{jk}N_{jk,c'}^{|g,0\rangle}}
\quad\mathrm{and}\quad 
p_{|{g,1\rangle}} = \frac{\sum_{j}N_{jb,c'}^{|g,1\rangle}}{\sum_{jk}N_{jk,c'}^{|g,1\rangle}}.
\end{equation}
We correct $N_{jk}^s$ for image 2 detection infidelity $\epsilon_\mathrm{inf}$ and loss to anti-trapped ${}^3\mathrm{P}_2$ occuring during that image $\epsilon_\mathrm{i,loss}$ via
\begin{equation}
N_{jb, c'}^s = (1 + \epsilon_\mathrm{inf} + \epsilon_\mathrm{i,loss}) N_{jb}^s,
\end{equation}
such that for experiments including the repumping step, $p_{|{g,0\rangle}}$ and $p_{|{g,1\rangle}}$ represent atom populations remaining in the trapping potential at the end of the reset sequence. When calculated for experiments without the repumping step, $p_{|{g,0\rangle}}$ and $p_{|{g,1\rangle}}$ are reduced by the populations trapped in $m$ at the end of the reset sequence. The results are summarized in table~\ref{tab:ndsd}.

\subsection{Reset through global control vs. feed-forward}

The global approach, with a repumping step at the end of the sequence, re-initializes the largest fraction of atoms in $g$. We measure $p_{|g,0\rangle} = \UpNDprob{}$ and $p_{|g,1\rangle} = \DownNDprob{}$ for this method, with the losses dominated by vacuum lifetime and Raman scattering to anti-trapped ${}^3\mathrm{P}_2$. When we assess the atom return to $g$ without the addition of the repump, we find that the global approach leaves $0.6(4)\%$ and $1.3(4)\%$ of $|g,0\rangle$ and $|g,1\rangle$ atoms in $m$ at the end of the sequence, respectively. For $\up$, this arises from a shelving error; while for $|g,1\rangle$, the error stems from optical pumping and clock polarization impurities. The latter two error sources are absent in the feed-forward approach---indeed, we find that the $|g,1\rangle$ population trapped in $m$ for this reset scheme is smaller. However, the feed-forward reset suffers from its own errors associated with correctly identifying $|g,0\rangle$ and $|g,1\rangle$ states in the first image as well as additional Raman scattering events that occur during image-processing time. As such, the feed-forward method ends up with $2.8(5)\%$ and $0.7(1)\%$ of $|g,0\rangle$ and $|g,1\rangle$ atoms trapped in $m$ (the latter number is inferred).

\begin{table*}%The best place to locate the table environment is directly after its first reference in text
\caption{\label{tab:ndsd}%
Error budget for detection fidelities $\mathcal{P}$ and ground-state re-initialization probabilities p in a clock-based non-destructive state-detection protocol. For each error, we indicate whether the physical mechanism responsible for the error is measured (m) or calculated (c). 
}
\begin{ruledtabular}
\renewcommand{\arraystretch}{1.1}
\begin{flushleft}
{State-detection}
\end{flushleft}\vspace*{-0.9\baselineskip}
\begin{tabular}{lllll}
{(\%)} & \multicolumn{2}{c}{$|g,0\rangle$} & \multicolumn{2}{c}{$|g,1\rangle$}\\
\colrule
\bf{$\mathcal{P}$ measured} & \multicolumn{2}{c}{\bf{98.6(2)}} & \multicolumn{2}{c}{\bf{99.4(1)}}\\
\colrule
Error   & current & projected & current & projected \\
\colrule
{Image infidelity(m)} &
0.23(6) & 0.23 & 0.18(9) & 0.18\\
{$\mathrm{{}^3P_0\rightarrow{}^3P_1 (m,c)}$}&
0.7(1) & 0.5 & -- & --\\
{Natural lifetime of $\mathrm{{}^3P_0}$ (c)} &
0.10(1) & 0.1 & -- & --\\
{Shelving error (m)} &
0.28(3) & 0.1& -- & --\\
{$\sigma^{-}$ clock polarization impurity (m)} &
-- & -- & 0.5(2) &  0.05\\
\bf{Total estimate} &
\bf{1.3(1)} & \bf{0.93} & \bf{0.7(2)} & \bf{0.23}\\
\end{tabular}
\begin{flushleft}
{Re-initialization in $g$}
\end{flushleft}\vspace*{-0.9\baselineskip}
\begin{tabular}{lllllll}
{(\%)} & \multicolumn{4}{c}{global control} & \multicolumn{2}{c}{local feed-forward} \\
{} & \multicolumn{2}{c}{$|g,0\rangle$} & \multicolumn{2}{c}{$|g,1\rangle$} & $|g,0\rangle$ & $|g,1\rangle$ \\
\colrule
\bf{$p$ measured (with repump)} &
\multicolumn{2}{c}{\bf{97.4(3)}} & \multicolumn{2}{c}{\bf{99.0(2)}} & \bf{97.6(3)} & --\\
\bf{$p$ measured (without repump)} &
\multicolumn{2}{c}{\bf{96.8(3)}} & \multicolumn{2}{c}{\bf{97.7(3)}} & \bf{94.7(5)} & \bf{98.4(3)}\\
\colrule
Error source & current & projected & current & projected & \multicolumn{2}{c}{current}\\
\colrule
Atom loss\\
{Vacuum loss (m)} &
0.8(2) & -- & 0.8(2) & -- & 0.8(2) & 0.8(2)\\
{$\mathrm{{}^3P_1\rightarrow{}^3P_2 (m,c)}$}  & -- & -- & 0.19(2) & 0.1 & -- & 0.19(2)\\
{$\mathrm{{}^3P_0\rightarrow{}^3P_2 (m,c)}$} &
1.3(2) & 0.8 & -- & -- & 1.6(2) & -- \\
{Shelving heating (m)} &
0.1 & 0.1 & -- &-- & 0.1 & -- \\
\bf{Total estimate} &
\bf{2.2(3)} & \bf{0.9} &\bf{1.0(2)} &   \bf{0.1} & \bf{2.5(3)} & \bf{1.0(2)}\\
\colrule
Residual population in $m$\\
Optical pumping(m) & -- & -- & 0.28(3) & 0.03 & -- & --\\
$\sigma^{-}$ clock polarization impurity (m) & -- & -- & 1.0(4) & 0.1 & -- & --\\
Shelving error (m) & 0.56(6) & 0.2 & -- & -- & 0.56(6) & --\\
Clock suppression error (m) & -- & -- & -- & -- & -- & 0.1\\
Image 1 detection infidelity (m) & -- & -- & -- & -- &  1.4(2) & 0.6(1)\\
{$\mathrm{{}^3P_0\rightarrow{}^3P_1 (m,c)}$\footnote{\label{foot:ndsd}When the scattering event or decay occurs between image 1 and feed-forward operation, the atom is shelved to $m$.}} & -- & -- & -- & -- & 1.9(3) & --\\
{Natural lifetime of $\mathrm{{}^3P_0}$ (c)${}^\mathrm{\ref{foot:ndsd}}$} &
-- & -- & -- & -- & 0.44(3) & --\\
\bf{Total estimate} & \bf{0.56(6)} & \bf{0.2} &\bf{1.3(4)} &   \bf{0.13} & \bf{4.3(4)} & \bf{0.7(1)}\\
\end{tabular}
\end{ruledtabular}
\end{table*}

\begin{table*}%The best place to locate the table environment is directly after its first reference in text
\caption{\label{tab:mco}%
Contrast loss budget for mid-circuit measurement (MCM) and mid-circuit reset. For each error, we indicate whether the physical mechanism responsible for the error is measured (m) or calculated (c). 
}
\begin{ruledtabular}
\renewcommand{\arraystretch}{1.1}
\begin{tabular}{lllllll}
(\%) & \multicolumn{2}{l}{\textrm{Ground-state MCM}}&
\multicolumn{2}{l}{\textrm{Mid-circuit reset}} & \multicolumn{2}{l}{\textrm{Metastable-state MCM}}\\
&
\textrm{Ancilla}&
\textrm{Data}&
\textrm{Ancilla}&
\textrm{Data}&
\textrm{Ancilla}&
\textrm{Data}\\
\colrule
\bf{$\mathcal{C}$ measured} & \bf{98.2(6)} & \bf{95.5(1.0)} & \bf{97.7(5)} & \bf{95.2(8)} & \bf{96.2(8)} & \bf{90(1)}\\
\colrule
Error estimate &\\
\colrule
{SPAM}\\
{Vacuum loss (m)} &
0.4(2) & 1.0(2) & 0.8(2) & 0.9(2) & 0.4(2) & 0.9(5)\\
{Optical pumping (m)} &
0.6(1) & 0.6(1) & 0.6(1) & 0.6(1) & 0.28(5) & 0.28(5)\\
{Image infidelity (m)} &
-- & 0.2(1) & 0.5(2) & 0.5(2) & -- & 0.4(2)\\
{Push-out (m)} &
-- & 0.09(1) & 0.09(1) & 0.09(1) & -- & --\\
{$\mathrm{{}^3P_1\rightarrow{}^3P_2}$ (m,c)} & -- & 0.07(1) & 0.08(1) & 0.08(1) & -- & 0.08(1)\\
{$\mathrm{{}^3P_0\rightarrow{}^3P_1}$ (m,c)} &
-- & -- & -- & -- & 0.06(1) & 0.9(1)\\
{$\mathrm{{}^3P_0\rightarrow{}^3P_2}$ (m,c)} &
-- & -- & -- & -- & 0.03(1) & 0.09(1) \\
{Natural lifetime of $\mathrm{{}^3P_0}$ (c)} &
-- & -- & -- & -- & 0.03(1) & 0.38(3)\\
{$m$ single-qubit gate error (m)} &
-- & -- & -- & -- & 0.75(3) & 1.0(4)\\
{Shelving error (m)} & -- & -- & -- & -- & 0.28(3) & 0.56(5)\\
{Shelving heating (m)} &
-- & -- & -- & -- & 0.05 & 0.1\\
{$\sigma^{-}$ clock polarization impurity (m)} &
-- & -- & -- & -- & -- & 0.5(2)\\
\bf{SPAM error} &
\bf{\textrm{1.0(2)}} & \bf{\textrm{2.0(2)}}  & \bf{\textrm{2.1(3)}} & \bf{2.2(3)} & \bf{1.9(2)} & \bf{5.2(7)}\\
\colrule
{Procedure}\\
{Vacuum loss (m)} &
0.04(2)  & 0.10(2) &  0.15(3) &  0.08(2) & 0.04(2) & 0.14(9)\\
{Image infidelity (m)} & 
0.5(3) & -- & -- & -- & 0.4(2) & --\\
{Push-out (m)} & 
0.09(1) & -- & -- & -- & -- & -- \\
{$\mathrm{{}^3P_1\rightarrow{}^3P_2}$ (m,c)} & 0.07(1) & -- & -- & -- & 0.07(1) & --\\
{$\mathrm{{}^3P_0\rightarrow{}^3P_1}$ (m,c)} &
-- & 1.2(2) & -- & 1.2(2) & 0.44(7) & 1.1(2)\\
{$\mathrm{{}^3P_0\rightarrow{}^3P_2}$ (m,c)} &
-- & 0.7(1) & -- & 0.7(1) & 0.02(1) & 0.8(1) \\
{Natural lifetime of $\mathrm{{}^3P_0}$ (c)} &
-- & 0.010(1) & -- & 0.09(1) & 0.07(1) & 0.15(1)\\
{Shelving error (m)} &
-- & 0.56(5) & -- & 0.56(5) & 0.28(3) & --\\
{Shelving heating (m)} &
-- & 0.1 & -- & 0.1 & 0.05 & --\\
{Clock suppression error (m)} &
0.1 & -- & 0.1 & -- & -- & 0.1\\
{$\sigma^{-}$ clock polarization impurity (m)} &
-- & -- & -- & -- & 0.5(2) & --\\
\bf{Procedure error} &
\bf{\textrm{0.8(3)}} & \bf{\textrm{2.8(2)}}  & \bf{\textrm{0.25(3)}} & \bf{2.7(2)} & \bf{1.9(3)} & \bf{2.3(2)}\\
\colrule
\bf{Total estimate} &
\bf{\textrm{1.8(4)}} & \bf{\textrm{4.7(3)}}  & \bf{\textrm{2.4(3)}} & \bf{\textrm{4.9(4)}} & \bf{\textrm{3.8(4)}} & \bf{\textrm{7.5(8)}}\\
\end{tabular}
\end{ruledtabular}
\end{table*}

\section{\label{sec:budget}Error budget}

To check whether we understand all errors and losses present in the non-destructive state-detection and in mid-circuit operations, we measure and analyze each known error mechanism independently and estimate its effect on the quantity of interest. The error budgets  are presented in Tables~\ref{tab:ndsd}~and~\ref{tab:mco}. Below we summarize the procedures used to determine the quoted numbers.

\textbf{Image infidelity} is extracted from a separate \enquote{calibration} data set, where all operations except magnetic field and tweezer depth ramps are removed. These are used to compute the probabilities of misidentifying a bright event as a dark one and a dark event as a bright one.

\textbf{${}^3P_1$$\rightarrow$${}^3P_2$ Raman scattering} This is a Raman process where 759 nm tweezer photons scatter from the intermediate ${}^3S_1$ state, changing the state of the atom from ${}^3P_1$ to ${}^3P_2$. Since ${}^3P_2$ is anti-trapped at 759 nm, the atoms are subsequently lost from the traps. With repeated imaging, we measure this loss to be $0.19(2)\%$ per image (fig.~\ref{fig:magicangle}). This is near to our \textit{ab initio} calculations of scattering rates, given known matrix elements. %We note that this error source contributes to both atom loss and to discrimination errors between the bright $b$ and dark $d$ events. For the former, unless a there is an additional image present in the experimental sequence, the loss is absorbed into the measured vacuum loss (see below). For the latter, the error will map $b$ to $d$ only if the scattering event occurs before the atom scatters enough photons to cross the photon count threshold. We take this effect into account. 

\textbf{${}^3P_0$$\rightarrow$${}^3P_1$ Raman scattering.} This is a Raman process caused by 759 nm tweezer photons, which changes the state of the atom from ${}^3P_0$ to ${}^3P_1$. Since ${}^3P_1$ is short-lived, the atoms subsequently end up in $g$. We extract the scattering rate of this process $\Gamma_{{}^3P_0\rightarrow{}^3P_1}$ by initializing atoms in $m$, recording the $g$ population as a function of wait time $t$ and fitting with $1-\exp(-\Gamma_{{}^3P_0\rightarrow{}^3P_1}t)$ model. We repeat this measurement for a range of trap depths $U$ obtaining a linear relation between $\Gamma_{{}^3P_0\rightarrow{}^3P_1}$ and $U$. The error is then computed as the total probability of the scattering event occurring, given the time atom in $m$ spends in the trap depth $U$. We note that this error maps a $d$ event to $b$, and thus will contribute only if the atom can scatter enough photons during imaging to cross the photon count threshold. We take this effect into account. 

\textbf{${}^3P_0$$\rightarrow$${}^3P_2$ Raman scattering.} This is a Raman process mediated by 759 nm photons, changing the state of the atom from ${}^3P_0$ to ${}^3P_2$. Since ${}^3P_2$ is anti-trapped at 759 nm, the atoms are subsequently lost from the traps. With \textit{ab initio} calculations involving known matrix elements, we calculate the ratios of  $\Gamma_{{}^3P_0\rightarrow{}^3P_2}$ and $\Gamma_{{}^3P_0\rightarrow{}^3P_1}$ scattering rates and with the $\Gamma_{{}^3P_0\rightarrow{}^3P_1}$ values measured as a function of trap depth $U$ (see above), we compute corresponding $\Gamma_{{}^3P_0\rightarrow{}^3P_2}$. The error is then calculated as the probability of a scattering event occurring, given the time atom in $m$ spends in the trap depth $U$. 

\textbf{Natural lifetime of $\mathrm{{}^3P_0}$.} Corresponding error calculated using measured natural lifetime from~\cite{Xu2014Measurement}.  

\textbf{Vacuum loss} is measured for each experiment in a separate \enquote{calibration} experiment, where all operations except magnetic field and tweezer depth ramps are removed. We correct the results for image infidelity. The observed error is consistent with independently determined vacuum lifetime, in addition to the imaging loss we expect. %, we note that the measurement will also contain a small contribution from ${}^3P_1$$\rightarrow$${}^3P_2$ Raman scattering process (see above) during the last image. 

\textbf{Optical pumping.} We characterize the optical pumping performance directly by preparing atoms in $\ket{g,0}$ and subsequently applying a push-out operation. The error is taken as the residual population fraction, corrected for the image infidelity. We note that for the ground-state mid-circuit operations, this state-preparation error reduces the contrast by $2\epsilon_{OP}$.

\textbf{Push-out} operation is realized by resonantly driving $\ket{g,0}\leftrightarrow\ket{{}^3\mathrm{P}_1, F'=3/2, m_{F'}=+3/2}$ transition for a certain time $t$. During this procedure, the off-resonant scattering on the $\ket{g,1}\leftrightarrow\ket{{}^3\mathrm{P}_1, F'=3/2, m_{F'}=+1/2}$ transition leads to $\ket{g,1}\rightarrow\ket{g,0}$ pumping process, where the pumped atoms are immediately ejected from the traps. Initializing atoms in $\ket{g,1}$ and recording atom survival as a function of push-out beam pulse length, we measure the off-resonant scattering rate $\tau$ and compute the push-out error as $1-\exp(-t/\tau)$.

\textbf{Shelving error} is the fraction of the population remaining in the original state after $S_1$ or $S_0$ operation, defined in the main text as $\epsilon_S$. We measure it by repeating the shelving operation odd number of times and detecting population that remains in $g$. The results are presented in fig.~\ref{fig:shelve}(c). %We note that if the shelving operation is performed multiple times, the error accumulates accordingly. 

\textbf{Shelving heating} is measured by repeating $S_1$ or $S_0$ operation even number of times, followed by repumping any residual population back to $g$. We observe a quadratic relation between the number of shelvings and atom loss. The quoted shelving heating is extracted from the fit to the data.

\textbf{Clock suppression error} is the population fraction shelved by $S_1$ or $S_0$ operation, in spite of the 532-nm-induced light-shift. The measurement of this error is presented in fig.~\ref{fig:clock}(b).

\textbf{$\sigma^-$ clock polarization impurity.} We measure the error in $S_1$ shelving operation due to polarization impurities present in the $\sigma^-$ clock drive. We initialize atoms in $\ket{g,1}$ and apply a single $S_1$ operation. After accounting for state-preparation errors, we observe that a fraction of atoms is shelved to $m$, which we attribute to a $\sigma^+$ polarization component present in the drive. %With two $S_1$ operations, this shelved fraction doubles, suggesting a $\sigma^+$ polarization component present in the drive. We confirm those atoms are indeed shelved, and not lost from the traps, by repeating the experiment with an additional repumping step at the end of the sequence.

\textbf{$m$ single-qubit error.} The error in the $m$ single-qubit gates is extracted from randomized benchmarking measurements (fig.~\ref{fig:control}(a)), and scaled by the time the pulse is applied in the relevant experiment. %The error in a single $X$ gate is taken as 2/7 of the error per Clifford gate, while the contrast loss during continuous Rabi oscillations is assumed to be proportional to the time the $m$ qubit drive is applied for. 

\section{\label{sec:timing_diagram} Timing diagram of mid-circuit operations}

\begin{figure}
\centering
\includegraphics{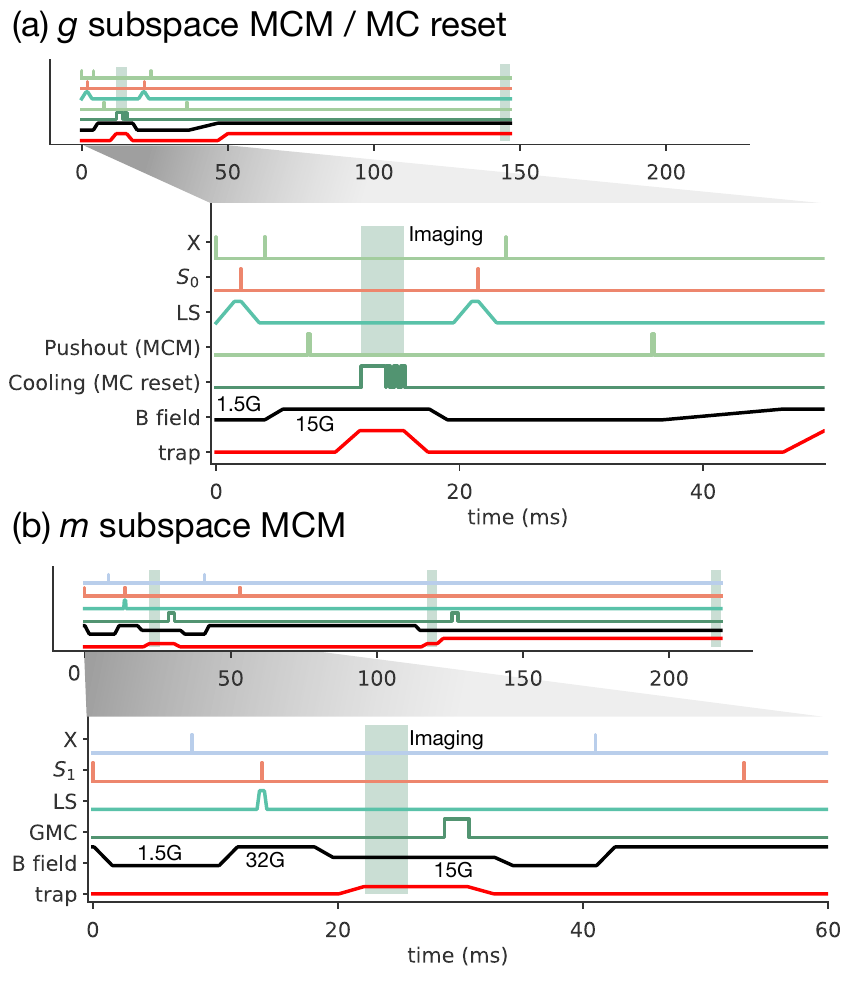}
\caption{\label{fig:sequence} Experimental sequence of the MCOs, with the timing is counted after the initial imaging and state preparation. For MCM, the freen shaded region indicates imaging. (a) Sequence for the ground-state MCM and MC reset. Both share a similar experimental sequence, except for the spin detection, cooling and spin reset. (b) Sequence for the metastable MCM. Here,an extra image is taken at the end to asses atom loss.}
\end{figure}

The detailed experimental sequence used in fig.~\ref{fig:mco} is given in the fig.~\ref{fig:sequence}. Below we highlight differences between the ground- and metastable-qubit operations. We note that in both cases there are substantial $\approx100$ ms delays between images associated with the charge clearing on the EMCCDs required for high signal/noise images; this will be avoidable in the future. The mid-circuit operations on their own take about $20~$ms, including shelving, trap ramps, field ramps, the image or reset, and data qubit unshelving.

For the ground-state mid-circuit measurement and mid-circuit reset experiments, the light shift beam (discussed in Appendix~\ref{subsec:LS}) is ramped up for 1.5 ms. For the imaging/cooling and push-out spin detection, we ramp up the magnetic field to 16 G, and increase the trap depth to 0.4 mK.

For the metastable qubit mid-circuit measurement, we used a 400 $\upmu$s-long pulse with a smooth envelop for the light shifting ($L$) beam to minimize the time of the $L$ beam while suppressing the loss from the procedure. For the $S_1$ operation, we increase the magnetic field to 32 G from the 1.5 G to reduce errors from polarization impurity. The imaging in this experiment was operated in the depth of $\approx$0.2 mK, which is close to half of the depth of the typical imaging operation, to reduce Raman scattering errors on the data qubit. 

\section{\label{subsec:QPT}Quantum Process Tomography for ground-state mid-circuit measurement}

In order to identify error sources and benchmark the fidelity of the mid-circuit measurement (MCM), we perform quantum process tomography \cite{chuang1997prescription, jevzek2003quantum, schindler2013quantum, chow2012universal}. We prepare the ancilla and data qubits in a set of input states, $\ket{g,0}$, $\ket{g,1}$, $\ket{g,0}$, $(\ket{g,0}+\ket{g,1})/\sqrt{2}$, and $(\ket{g,0}-i\ket{g,1})/\sqrt{2}$, feed these states in to the MCM operation, and perform state tomography on the resulting output states. In our destructive measurement protocol, the detection of no atom corresponds to events in which the atom was in the $\ket{g,0}$ state, but also to events in which the atom was lost from the trap or ended up in $\ket{m,0}$ or $\ket{m,1}$ states. To separate the probabilities of measuring $\ket{g,1}$ and $\ket{g,0}$ from the probabilities that the atom is lost or in the metastable state, we also perform the MCM without the destructive measurement pulses. We use the resulting atom detection probabilities to normalize the $\ket{g,1}$ and $\ket{g,0}$ detection probabilities used in state tomography. Implicit in this procedure is the assumption that the trap loss and metastable state shelving errors are independent of the nuclear spin state. Using these normalized detection probabilities, the process we reconstruct describes the MCM for events without loss or shelving errors. The output density matrices from this process can then be scaled by the probability of no loss and no shelving errors \cite{bhandari2016general}.

\begin{figure}
\centering
\includegraphics{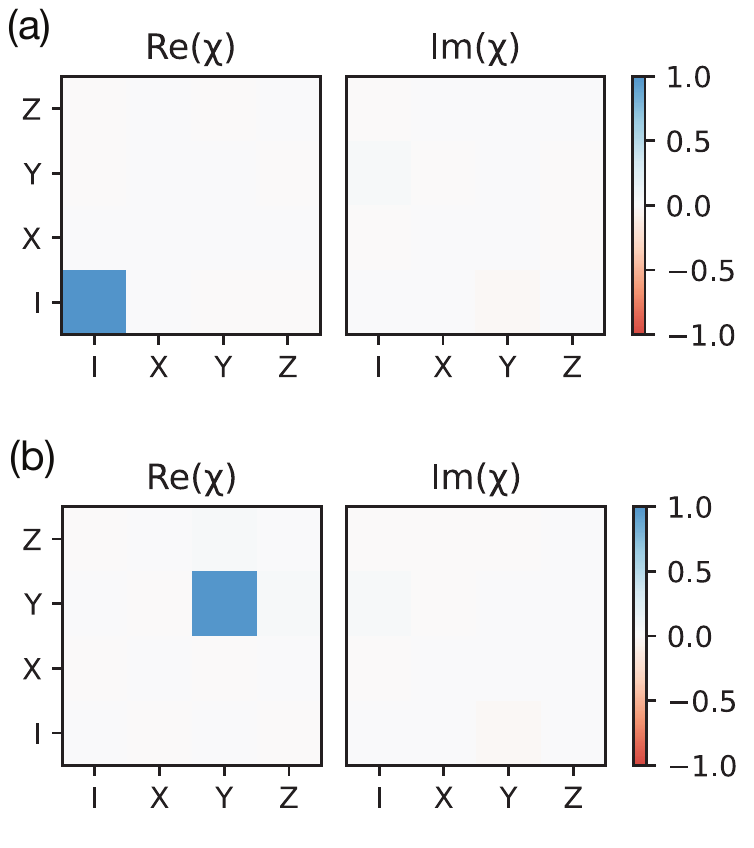}
\caption{\label{fig:qpt} Quantum process tomography. Reconstructed $\chi$ process matrices for (a) ancilla qubits and (b) data qubits. Here, $\{I,X,Y,Z\}$ denote Pauli matrices, i.e. $\{I,R_X(\pi),R_Y(\pi),R_Z(\pi)\}$ rotations.}
\end{figure}

We use an iterative maximum likelihood estimation algorithm to reconstruct the (lossless) MCM process, constraining the process to be completely positive and trace preserving \cite{jevzek2003quantum, hradil20043, lvovsky2004iterative}. Despite being lossless, this process still describes the dephasing, depolarization, and unitary rotation errors present in our MCM operation. The Choi matrices obtained from the reconstruction algorithm are converted to $\chi$ matrices (fig.~\ref{fig:qpt}) and used to calculate the process fidelities as $F_p=Tr(\chi \chi_{ideal})$ where $\chi_{ideal}$ describes the ideal MCM process \cite{chuang1997prescription}. Due to the field splitting, there is phase accumulation between the qubit states during the MCM. For the ancilla qubits, we calibrate the time between initialization and readout so that $\chi_{ideal}$ represents the identity. For the data qubits, we find that a qubit echo pulse minimizes the dephasing errors that are present on the timescales required to collect all data. The resulting process for the data qubits has a rotation given by $R_Z(\theta_1)R_X(\pi)R_ Z(\theta_2)$, where $\theta_1$ and $\theta_2$ are chosen so that the full rotation of the MCM process is equivalent to $R_Y(\pi)$. We find process fidelities for the data and measurement qubits $F_{p,\mathrm{DQ}} = 0.972(5)$ and $F_{p,\mathrm{AQ}} = 0.979(6)$ respectively. These process fidelities can be converted to average state fidelities as $F_{av}=(d F_p+1)/(d+1)$ \cite{nielsen2002simple, chow2012universal}. In our case $d=2$, yielding $F_{av,\mathrm{DQ}} = 0.981(4)$ and $F_{av,\mathrm{AQ}} = 0.986(4)$. Including a scaling factor in the output density matrices that accounts for loss and shelving errors results in average state fidelities $F_{av,\mathrm{DQ}} = 0.961(3)$ and $F_{av,\mathrm{AQ}} = 0.972(4)$. Using the reconstructed process, we can also directly calculate the output state fidelities, including loss and shelving errors, for the specific input state $(\ket{g,0}-i\ket{g,1})/\sqrt{2}$. This gives  $F_{\ket{-y},\mathrm{DQ}} = 0.948$ and $F_{\ket{-y},\mathrm{AQ}} = 0.978$, which can be compared to the Ramsey contrast of ancilla and data qubits in Fig. 5, $\mathcal{C}_\mathrm{MCM}^{\mathrm{DQ}} = \MCMDQcontrast$ and $\mathcal{C}_\mathrm{MCM}^{\mathrm{AQ}} = \MCMAQcontrast$.

\bibliography{theBib}% Produces the bibliography via BibTeX.

\end{document}